\documentclass[english]{revtex4}
\usepackage[T1]{fontenc}
\usepackage[latin1]{inputenc}
\usepackage{graphicx}
\usepackage{amssymb}

\makeatletter

\providecommand{\tabularnewline}{\\}

\usepackage{bm}
\usepackage{psfrag}
\usepackage{hyperref}

\usepackage{babel}
\makeatother
\begin{document}

\title{Phase space and quark mass effects in neutrino emissions in a color
superconductor}

\author{Qun Wang}

\affiliation{Department of Modern Physics, University of Science and Technology
of China, Anhui 230026, People's Republic of China}

\author{Zhi-gang Wang}

\thanks{On leave from Department of Physics, North China Electric Power University,
Baoding 071003, P. R. China}

\affiliation{Department of Modern Physics, University of Science and Technology
of China, Anhui 230026, People's Republic of China}

\author{Jian Wu}

\affiliation{Department of Modern Physics, University of Science and Technology
of China, Anhui 230026, People's Republic of China}

\begin{abstract}
We study the phase space for neutrino emissions with massive quarks
in direct Urca processes in normal and color superconducting quark
matter. We derive in QCD and the NJL model the Fermi momentum reduction
resulting from Fermi liquid properties which opens up the phase space
for neutrino emissions. The relation between the Fermi momentum and
chemical potential is found to be $p_{F}\approx\mu(1-\kappa)$ with
$\kappa$ depending on coupling constants. We find in the weak coupling
regime that $\kappa$ is a monotonously increasing function of the
chemical potential. This implies quenched phase space for neutrino
emissions at low baryon densities. We calculate neutrino emissivities
with massive quarks in a spin-one color superconductor. The quark
mass corrections are found to be of the same order as the contributions
in the massless case, which will bring sizable effects on the cooling
behavior of compact stars. 
\end{abstract}
\maketitle

\section{Introduction and Conclusion}

The baryon density in the core of a compact star is likely to reach
several times the nuclear saturation density, $\rho_{0}\sim0.16\;\mathrm{fm}^{-3}$
or $2.7\times10^{14}\mathrm{g}/\mathrm{cm}^{3}$. At such a high density,
nucleons in nuclear matter are crushed into their constituents, i.e.
quarks and gluons. Itoh proposed a model for quark stars in 1970 \cite{Itoh:1970uw}.
The deconfinement transition to quark matter was suggested by Collins
and Perry in 1975 \cite{Collins:1974ky} based on the asymptotic freedom
in quantum chromodynamics. In the same paper they also mentioned the
possibility that quark matter could be a superfluid or a superconductor
resulting from the attractive interquark force in some channels. Barrois,
Bailin and Love developed this novel idea and studied the unusual
variant of superconductivity in quark matter, which we now call color
superconductivity \cite{Barrois:1977xd,Bailin:1983bm}. They did their
calculations in the framework of weak coupling approach and did not
take into account the dynamic screening of magnetic gluons which are
dominant agents in pairing quarks. Therefore the gap or equivalently
the transition temperature they obtained are too small to be of relevance
to any sizable observables. About fifteen years later the color superconductivity
had been re-discovered by several groups who found the gap could be
large enough to bring some real effects in compact stars \cite{Alford:1997zt,Alford:1998mk,Rapp:1997zu,Son:1998uk,Pisarski:1999tv,Hong:1999fh}.
{[}For recent reviews on color superconductivity, see, for example,
\cite{Rajagopal:2000wf,Alford:2001dt,Reddy:2002ri,Casalbuoni:2003wh,Schafer:2003vz,Rischke:2003mt,Buballa:2003qv,Ren:2004nn,Shovkovy:2004me}.{]}

At asymptotically high baryon density or chemical potential, the color-flavor
locked (CFL) phase is favored \cite{Alford:1998mk}. The rigorous
weak coupling approach of QCD \cite{Son:1998uk,Pisarski:1999tv,Brown:1999aq,Wang:2001aq,Schmitt:2002sc,Ipp:2003cj,Reuter:2004kk,Gerhold:2005uu,Noronha:2006cz}
is justified. {[}For recent reviews on the QCD weak coupling approach
in normal dense quark matter or in color superconductivity, see \cite{Rischke:2003mt,Ren:2004nn,Wang:2004pa,Rebhan:2005hw,Gerhold:2005fi,Gerhold:2005nt,Reuter:2006iw}.{]}
The situation is more complicated at lower densities realized in compact
stars. In this case, the strange quark mass $m_{s}$ is in the range
from its current mass $\sim$100 MeV to its constituent mass $\sim$500
MeV. The main effect of $m_{s}$ is to cause a mismatch in chemical
potentials between strange quarks and light quarks. The CFL phase
is therefore broken down to less symmetric phases. Moreover the $\beta$
equilibrium and electric and color charge neutrality also give rise
to a mismatch between the Fermi momenta of the quarks that form Cooper
pairs \cite{Iida:2000ha,Alford:2002kj,Steiner:2002gx,Huang:2002zd,Neumann:2002jm,Ruster:2003zh}.
If the mismatch in chemical potentials for pairing quarks is large
enough, the conventional BCS pairing becomes questionable \cite{Huang:2004bg,Casalbuoni:2004tb,Giannakis:2004pf,Alford:2005qw,Fukushima:2005cm,Rajagopal:2005dg}
and the true ground state of quark matter in compact stars has to
include different, unconventional superconducting states \cite{Giannakis:2005vw,Gorbar:2005rx,Schafer:2005ym,Kitazawa:2006zp,He:2006tn}. 

Single-flavor Cooper pairing is the simplest option for neutral quark
matter. Contrary to other unconventional pairing mechanisms, it is
allowed for arbitrarily large mismatches between the Fermi momenta
of different quark flavors. Single-flavor pairing in the color anti-triplet
channel is possible only in the symmetric spin-one channel \cite{Pisarski:1999tv,Schafer:2000tw,Alford:2002rz,Schmitt:2002sc,Buballa:2002wy,Schmitt:2003xq,Schmitt:2004et,Aguilera:2005tg,Alford:2005yy},
because Pauli principle requires that the wave function of the Cooper
pair be antisymmetric with respect to the exchange of the pairing
quarks. This is in contrast to pairing of different flavors where
the antisymmetric spin-zero channel is allowed. 

Similar to superfluid Helium-3, where condensation occurs in spin
and angular momentum triplets \cite{Vollhardt}, the order parameter
in a spin-one color superconductor is a complex $3\times3$ matrix.
Therefore various possible phases emerge corresponding to different
patterns of the matrix order parameter. Among others, there are four
main spin-one color-superconducting phases: the color-spin locked
(CSL), planar, polar, and $A$ phases \cite{Schmitt:2002sc,Schmitt:2003xq,Schmitt:2004et}.
Except for the CSL phase, the gap functions of spin-one phases are
normally anisotropic in momentum space. The gap in the polar phase
vanishes at the south and north poles of the Fermi sphere, whereas
the gap in the planar phase is anisotropic but nonzero in any direction
of the quasi-particle momentum. The $A$ phase is special in the sense
that it has two gapped quasi-particle modes with different angular
structures, one of which has two point nodes. Spin-one color superconducting
phases play important roles in thermal properties in quark matter.
Spin-one phases have gaps of a few tens of KeVs, which can fit into
the cooling data of compact stars \cite{Popov:2005xa}. The anisotropies
especially the nodes in some spin-one phases significantly affect
various thermodynamical and transport properties like the specific
heat, the neutrino emissivity, the viscosity, the heat and electrical
conductivity, etc.. In Ref. \cite{Schmitt:2005wg}, we have computed
the neutrino emissivity from direct Urca (DU) processes and the specific
heat for the mentioned spin-one phases to obtain the resulting cooling
rates for compact stars. We have discussed the reason why the distribution
of neutrino emission from the $A$ phase breaks reflection symmetry
in space \cite{Schmitt:2005ee}. Similar calculations of neutrino
emissivity in DU processes in the 2SC phase can be found in Ref. \cite{Jaikumar:2005hy}. 

We know that DU processes in normal and color superconducting quark
matter is controlled by the phase space opened up through Fermi liquid
properties \cite{Iwamoto:1980eb}, which arises from the quark-quark
forward scattering amplitude. Up to now the amplitude has been derived
only via one gluon exchange \cite{Baym:1975va,Schafer:2004jp}, which
is a standard weak coupling perturbative approach with zero quark
mass. In this paper we derive the quark-quark scattering amplitude
via one gluon exchange but with massive quarks. We also do the same
in the NJL model. The forward scattering amplitude gives a general
formula $p_{F}=\mu(1-\kappa)$ for the relation between the Fermi
momentum $p_{F}$ and the chemical potential $\mu$, where $\kappa$
is the Fermi momentum reduction coefficient with respect to $\mu$
and depends on coupling constants. The phase space for DU processes
with massive quarks both in QCD and in the NJL model is then openned
up resulting from the Fermi momentum reduction. We find an interesting
behavior of $\kappa(\mu)$ as a function of $\mu$ in the weak coupling
regime that $\kappa(\mu)$ increases with increasing $\mu$ and approaches
its limit value as $\mu\rightarrow\infty$. It means that the phase
space for neutrino emissions is larger at high baryon densities than
at low ones. The property seems robust and independent of specific
models in computing quark-quark scattering amplitudes and Landau coefficients. 

In this paper we also extend our established formalism \cite{Schmitt:2005wg,Schmitt:2005ee}
by including quark masses to calculate neutrino emissivities in a
spin-one color superconductor. The quark masses bring much complexity
to quasi-particle dispersion relations in spin-one phases, as was
shown in Ref. \cite{Aguilera:2005tg,Aguilera:2006cj}. We use the
effective quark mass on the Fermi surface in the calculation, which
is determined by the quark-quark interaction in the Fermi liquid.
By assuming a natural form of the order parameters for spin-one phases,
we show that the quasi-particle dispersion relations bear the same
form as in the massless case, which much facilitates the calculation.
In the collinear and small mass limit, we derive neutrino emissivities
with massive quarks and identify mass corrections. The mass corrections
are of the same order as in the massless case. We find no mass correction
in the polar phase. The mass correction for the A phase approaches
zero at asymptotically low temperatures because all contributions
are from gapped modes. The mass corrections in the CSL and Planar
phases are found to be about 26\% and 15\% of the contributions in
the massless case at low temperatures respectively. This will have
sizable effects on the cooling of compact stars. These percentages
for the mass corrections are universal as long as $\kappa\ll1$, which
is fullfilled in QCD weak coupling approach and for realistic values
of coupling constants in the NJL model. Note that mass corrections
could be very different near the chiral phase transition or for strange
quarks with even larger quark masses. However our current formulation
is only valid at small mass limit with respect to the chemical potential.
For very large quark masses, the Landau Fermi liquid theory has to
be reformulated and the integral in the emissivity is very different
from the small mass limit. This is beyond the scope of the current
work. A future investigation of effects with large quark masses is
needed. 

Our convention for the metric tensor is $g^{\mu\nu}=\textrm{diag}(1,-1,-1,-1)$.
Our units are $\hbar=c=k_{B}=1$. Four-vectors are denoted by capital
letters, $K\equiv K^{\mu}=(k_{0},\mathbf{k})$, while $k=|\mathbf{k}|$.

\section{Phase space for direct Urca processes in quark matter and effective
quark mass }

\label{sec:phase-space}
\begin{figure}

\caption{\label{cap:phasespace}Phase space for DU processes with massless
quarks}

\includegraphics[%
  scale=0.8]{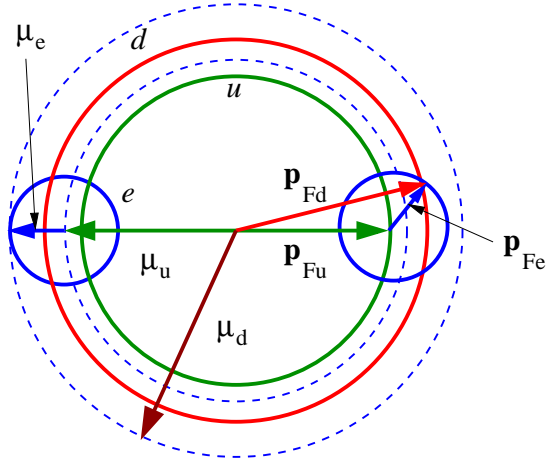}
\end{figure}

We study DU processes for light quarks $u+e^{-}\rightarrow d+\nu_{e}$
and $d\rightarrow u+e^{-}+\overline{\nu}_{e}$, where $e^{-}$ denotes
electrons and $\nu_{e}$ and $\overline{\nu}_{e}$ denote electron
neutrinos and anti-neutrinos respectively. The phase space is essential
for these processes to proceed. In normal quark matter, if quarks
are treated as free and massless, $\beta$ equilibrium requires $\mu_{d}=\mu_{u}+\mu_{e}$,
where $\mu_{d}$, $\mu_{u}$ and $\mu_{e}$ are chemical potentials
for $d$, $u$ quarks and electrons respectively. This leads to the
relation in Fermi momenta $p_{Fd}=p_{Fu}+p_{Fe}$. One sees that the
phase space is zero because there is no triangle inequality among
Fermi momenta. If the quark-quark interaction is switched on, Fermi
momenta are not the same as chemical potentials any more, instead
they get negative corrections from Landau Fermi liquid property \cite{Baym:1975va,Schafer:2004jp},
\begin{eqnarray}
p_{iF} & = & (1-\frac{C_{F}\alpha_{S}}{2\pi})\mu_{i}\,,\;\;\; i=u,d,\label{eq:pf-mu-qcd}\end{eqnarray}
where $\alpha_{S}$ is the strong coupling constant and $C_{F}=(N_{c}^{2}-1)/(2N_{c})$
is the eigenvalue of the Casimir operator of $SU_{c}(N_{c})$ group
in the fundamental representation with the number of colors $N_{c}=3$.
The corrections to Fermi momenta open up the phase space characterized
by the triangle inequality $p_{Fd}<p_{Fu}+p_{Fe}$. As illustrated
in Fig. \ref{cap:phasespace}, two dashed circles denote the Fermi
surfaces for free $d$ and $u$ quarks. They shrink to two smaller
solid circles after the interaction is turned on. The amount of reduction
in Fermi momentum for $d$ quarks is larger than that for $u$ quarks,
then there is a triangle among Fermi momenta implying non-vanishing
phase space. The corrections proportional to $\alpha_{s}$ is from
the quark-quark forward scattering via one gluon exchange with zero
quark mass. 

In this section we will re-analyze the phase space for DU processes
by deriving $p_{F}$ as a function of $\mu$ in QCD with quark masses
and some other features not considered in Ref. \cite{Baym:1975va,Schafer:2004jp}.
We also carry out the same task in the NJL model which has not been
done before. 

The relation between the Fermi momentum and the chemical potential
in the Landau Fermi liquid theory can be derived from \begin{eqnarray}
\frac{\partial\mu}{\partial n} & = & \frac{2\pi^{2}}{d_{g}\mu p_{F}}+f_{0}^{S}-\frac{1}{3}f_{1}^{S},\label{eq:dndmu}\end{eqnarray}
where $d_{g}=2N_{f}N_{c}$ is the quark degeneracy factor and $N_{f}$
the number of flavors. $n$ is the number density of quasi-particles.
The Landau coefficients $f_{0,1}^{S}$ are defined by \begin{eqnarray}
f_{l}^{S} & \equiv & \frac{2l+1}{4}\sum_{\sigma\sigma'}\int\frac{d\Omega}{4\pi}P_{l}(\cos\theta)f_{\mathbf{p}\sigma,\mathbf{p}'\sigma'}.\label{eq:landau-coef}\end{eqnarray}
Here $f_{\mathbf{p}\sigma,\mathbf{p}'\sigma'}$ is the Landau Fermi
interaction for quarks with spin indices $\sigma$ and $\sigma'$.
The angle $\theta$ is between $\mathbf{p}$ and $\mathbf{p}'$ both
taken on the Fermi surface. The integral is over all directions of
$\mathbf{p}$. The factor $1/4$ arises from the average over $\sigma$
and $\sigma'$. The factor $(2l+1)$ comes from the normalization
condition for Legendre polynomials $\int dxP_{l}^{2}(x)=2/(2l+1)$.
The density of states on the Fermi surface is given by \begin{equation}
N(0)\equiv\int d\tau\delta(E_{\mathbf{p}\sigma}-\mu)=\frac{d_{g}p_{F}^{2}}{2\pi^{2}}\left(\frac{\partial p}{\partial E_{\mathbf{p}\sigma}}\right)_{p_{F}},\label{eq:state-density}\end{equation}
where $(\partial p/\partial E_{\mathbf{p}\sigma})_{p_{F}}$ is the
inverse of the quasi-particle velocity on the Fermi surface. Note
that we have assumed that the flavor dependence of the quark mass
is weak and negligible, which is true for $\beta$ equilibrium where
$\mu_{u}\approx\mu_{d}$, so we suppressed the flavor indices of the
chemical potential $\mu$. 

\begin{figure}

\caption{\label{cap:k-mu-qcd}The Fermi momentum reduction coefficient $\kappa(\mu)$
as given in Eq. (\ref{eq:kappa-qcd}). The dotted line is the static
solution $\kappa=\frac{C_{F}\alpha_{S}}{2\pi}$. The solid line is
the branch of $\kappa(\mu)$ below the dotted line corresponding to
an initial value smaller than $\frac{C_{F}\alpha_{S}}{2\pi}$, while
the dashed line is that above it. (a) With the initial value $\kappa(\mu_{0}=0.3)=\frac{C_{F}\alpha_{S}}{2\pi}\pm0.08$;
(b) with $\kappa(\mu_{0}=0.3)=\frac{C_{F}\alpha_{S}}{2\pi}\pm0.1$.
Following Eq. (\ref{eq:exact-k}), the lower branch (solid line) is
the physical solution. }

\hspace{0.1cm}

\includegraphics[%
  scale=0.8]{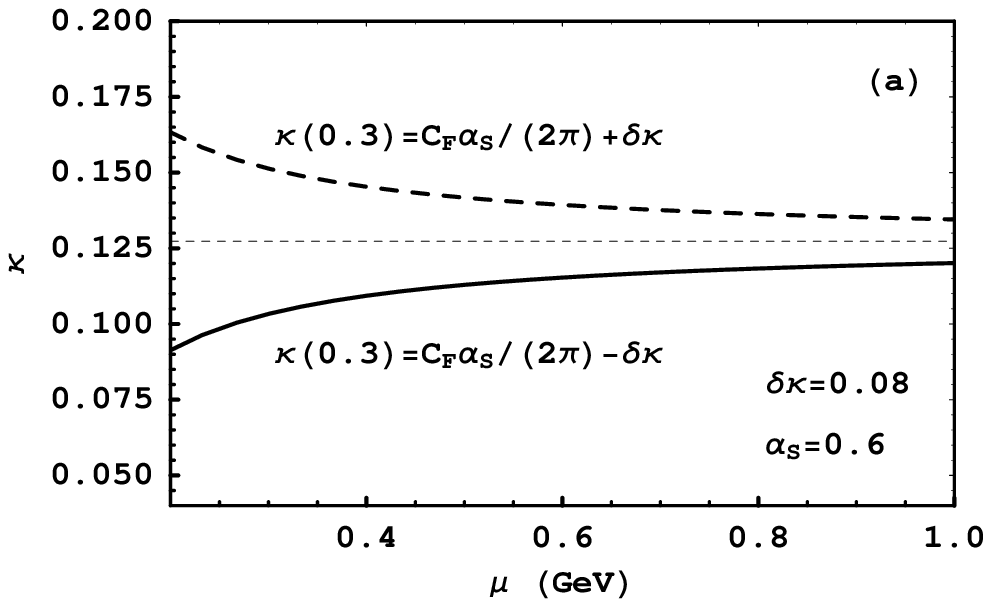}\hspace{0.5cm}\includegraphics[%
  scale=0.8]{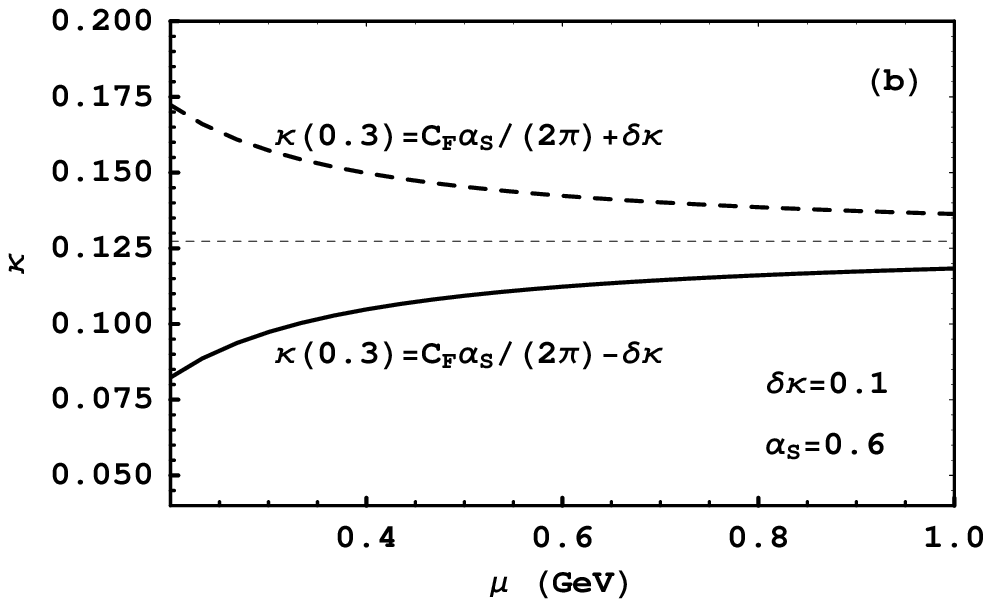}
\end{figure}

In QCD, the quark-quark scattering amplitude via one-gluon exchange
is given by \begin{eqnarray}
\mathcal{A}_{QCD} & = & \frac{C_{F}}{4N_{f}N_{c}}g^{2}\frac{1}{E_{\mathbf{p}_{1}}E_{\mathbf{p}_{2}}}\frac{4m^{2}-2P_{1}\cdot P_{2}}{(P_{1}-P_{2})^{2}}.\label{eq:fQCD}\end{eqnarray}
An average over colors, flavors and spins in the initial state and
a sum over all of them in the final state have been taken. Here $P_{1}=(E_{\mathbf{p}_{1}},\mathbf{p}_{1})$
and $P_{2}=(E_{\mathbf{p}_{2}},\mathbf{p}_{2})$ are on-shell 4-momenta
of two colliding quarks respectively, $m$ is the quark mass, and
$g$ is the coupling constant in QCD and $\alpha_{S}=g^{2}/(4\pi)$.
The amplitude diverges at small scattering angles $\theta$ since
$(P_{1}-P_{2})^{2}\approx-2p_{F}^{2}(1-\cos\theta)\rightarrow0$ if
$m\neq0$. The Landau coefficients can be obtained by applying Eq.
(\ref{eq:landau-coef}). Of course all Landau coefficients are divergent.
However the combination $f_{0}^{S}-\frac{1}{3}f_{1}^{S}$ is finite,
\begin{eqnarray*}
f_{0}^{S}-\frac{1}{3}f_{1}^{S} & = & \int\frac{d\Omega}{4\pi}\mathcal{A}_{QCD}(1-\cos\theta)\approx\frac{C_{F}}{4N_{f}N_{c}}g^{2}\frac{1}{\mu^{2}}\left(1-\frac{m^{2}}{p_{F}^{2}}\right).\end{eqnarray*}
Here the momenta and energies are set to the Fermi momentum and the
chemical potential respectively. Substituting the above expression
into Eq. (\ref{eq:dndmu}), one obtains \begin{equation}
\frac{d_{g}}{2\pi^{2}}p_{F}^{2}\frac{dp_{F}}{d\mu}=\left[\frac{2\pi^{2}}{d_{g}\mu p_{F}}+\frac{C_{F}g^{2}}{2d_{g}}\frac{1}{\mu^{2}}\left(2-\frac{\mu^{2}}{p_{F}^{2}}\right)\right]^{-1}\label{eq:dndmu2}\end{equation}
where we have used $n=\frac{d_{g}}{6\pi^{2}}p_{F}^{3}$. We assume
$\alpha_{S}$ is small. The Fermi momentum is related to the chemical
potential in the form \begin{equation}
p_{F}=\mu(1-\kappa),\label{eq:pf-mu}\end{equation}
where $\kappa$ is the Fermi momentum reduction coefficient with respect
to $\mu$ and is assumed to be a small dimensionless number. Then
the solution of Eq. (\ref{eq:dndmu2}) reads \begin{eqnarray}
\kappa(\mu) & = & \left[\kappa(\mu_{0})-\frac{C_{F}\alpha_{S}}{2\pi}\right]\frac{\mu_{0}^{2}}{\mu^{2}}+\frac{C_{F}\alpha_{S}}{2\pi},\label{eq:kappa-qcd}\end{eqnarray}
where we only kept the leading contribution. One can see in Eq. (\ref{eq:kappa-qcd})
the quark mass term belongs to the next-to-leading order and does
not appear. The Fermi momentum reduction coefficient $\kappa(\mu)$
in Eq. (\ref{eq:kappa-qcd}) is illustrated in Fig. \ref{cap:k-mu-qcd}.
There are two unconnected branches in $\kappa(\mu)$, one is above
the static solution $\kappa=\frac{C_{F}\alpha_{S}}{2\pi}$, the other
below it. For the upper branch, $\kappa(\mu)$ monotonously decreases
and approaches the limit vaule $\kappa=\frac{C_{F}\alpha_{S}}{2\pi}$
with increasing $\mu$. For the lower branch, the trend is opposite.
Both branches go towards the limit $\kappa=\frac{C_{F}\alpha_{S}}{2\pi}$
at asymptotically large $\mu$, i.e. $\kappa(\mu\rightarrow\infty)=\frac{C_{F}\alpha_{S}}{2\pi}$.
But at lower $\mu$ or lower baryon densities, there are two different
trends for $\kappa(\mu)$, one corresponds to more phase space and
the other to less phase space for neutrino emissions. 

If $\kappa$ is small and its dependence on $\mu$ is weak, Eq. (\ref{eq:dndmu2})
can be simplified to an algebraic equation without assuming $\alpha_{S}$
is small. The solution is \begin{eqnarray}
\kappa & = & \frac{C_{F}\alpha_{S}}{2\pi+5C_{F}\alpha_{S}}.\label{eq:exact-k}\end{eqnarray}
The above solution is lower than the limit value $\kappa(\mu\rightarrow\infty)=\frac{C_{F}\alpha_{S}}{2\pi}$
and can approach it when $\alpha_{S}\ll1$. It is interesting to compare
the solution (\ref{eq:kappa-qcd}) with (\ref{eq:exact-k}). If near
a specific value $\mu_{0}$ the dependence of $\kappa$ on the chemical
potential is negligible, $\kappa(\mu_{0})$ is given by Eq. (\ref{eq:exact-k})
satisfying $\kappa(\mu_{0})<\frac{C_{F}\alpha_{S}}{2\pi}$, which
leads to $\kappa(\mu)<\frac{C_{F}\alpha_{S}}{2\pi}$ for any values
of $\mu$ following Eq. (\ref{eq:kappa-qcd}). Thus the lower branch
of the solutions in Fig. \ref{cap:k-mu-qcd} is physical, implying
quenched phase space for neutrnio emissions at lower baryon densities. 

One can also determine from Eq. (\ref{eq:pf-mu}) and (\ref{eq:kappa-qcd})
the effective quark mass $m\sim\mu\sqrt{2\kappa}$ which is proportional
to the chemical potential. Note that the effective quark mass is determined
by the interquark interaction on the Fermi surface and can be much
larger than the current quark mass. 

The above results are from the QCD weak coupling approach and valid
at very high densities. At mediate densities realized in neutron stars,
one has to use more phenonmenological model like the NJL one. For
a review of the NJL model in quark matter, see, e.g., Ref. \cite{Buballa:2003qv}.
Now we consider the simplest case in the NJL model, a two flavor system,
whose Lagrangian is \begin{eqnarray}
\mathcal{L} & = & \overline{\psi}\left[i\gamma^{\mu}\partial_{\mu}-m\right]\psi+G_{S}\left[(\overline{\psi}\psi)^{2}+(\overline{\psi}i\gamma_{5}\bm{\tau}\psi)^{2}\right]-G_{V}(\overline{\psi}\gamma_{\mu}T^{a}\psi)^{2},\label{eq:lagragian}\end{eqnarray}
where $T^{a}=\lambda^{a}/2$ and $\lambda^{a}$ are the Gell-mann
matrices in color space. $\bm{\tau}=(\tau^{1},\tau^{2},\tau^{3})$
are Pauli matrices in flavor space. $G_{S}$ is the coupling constant
for the scalar and pseudoscalar channel, and $G_{V}$ is that for
the vector channel. In some sense, these two channels are independent
of each other. Hereafter we study each channel separately by setting
the coupling constant zero in the other channel. 

To obtain Landau coefficients $f_{0,1}^{S}$, we calculate the quark-quark
scattering amplitude, \begin{eqnarray}
\mathcal{A} & = & \mathcal{A}_{S}+\mathcal{A}_{PS}+\mathcal{A}_{V}=\frac{1}{N_{f}N_{c}}\frac{B_{1}P_{1}\cdot P_{2}-B_{2}m^{2}}{E_{\mathbf{p}_{1}}E_{\mathbf{p}_{2}}},\end{eqnarray}
 where $B_{1}$ and $B_{2}$ are two coefficients depending on coupling
constants $G_{S}$ and $G_{V}$, \begin{eqnarray}
B_{1} & = & 2G_{S}+C_{F}G_{V},\nonumber \\
B_{2} & = & G_{S}(2N_{c}N_{f}+1)+2C_{F}G_{V}.\end{eqnarray}
The amplitudes $\mathcal{A}_{S}$, $\mathcal{A}_{PS}$ and $\mathcal{A}_{V}$
for scalar, pseudoscalar and vector couplings respectively are given
by \begin{eqnarray}
\mathcal{A}_{S} & = & \frac{1}{2N_{f}N_{c}}G_{S}\frac{P_{1}\cdot P_{2}-(4N_{c}N_{f}-1)m^{2}}{E_{\mathbf{p}_{1}}E_{\mathbf{p}_{2}}},\nonumber \\
\mathcal{A}_{PS} & = & \frac{3}{2N_{f}N_{c}}G_{S}\frac{P_{1}\cdot P_{2}-m^{2}}{E_{\mathbf{p}_{1}}E_{\mathbf{p}_{2}}},\nonumber \\
\mathcal{A}_{V} & = & \frac{C_{F}}{N_{f}N_{c}}G_{V}\frac{P_{1}\cdot P_{2}-2m^{2}}{E_{\mathbf{p}_{1}}E_{\mathbf{p}_{2}}}.\end{eqnarray}
 An average over colors, flavors and spins in the initial state and
a sum over all of them in the final state have been taken. The Landau
coefficients can be obtained by using Eq. (\ref{eq:landau-coef}),
\begin{eqnarray}
f_{0} & = & \int\frac{d\Omega}{4\pi}\mathcal{A}=\frac{1}{N_{f}N_{c}}\left(B_{1}-B_{2}\frac{m^{2}}{\mu^{2}}\right),\nonumber \\
f_{1} & = & 3\int\frac{d\Omega}{4\pi}\mathcal{A}\cos\theta=-\frac{B_{1}}{N_{f}N_{c}}\frac{p_{F}^{2}}{\mu^{2}},\label{eq:landau-coef1}\end{eqnarray}
where the momenta and energies are set to the Fermi momentum and the
chemical potential respectively. Substituting $f_{0,1}^{S}$ of Eq.
(\ref{eq:landau-coef1}) into Eq. (\ref{eq:dndmu}), using $n=\frac{d_{g}}{6\pi^{2}}p_{F}^{3}$,
one obtains \begin{eqnarray}
\frac{d_{g}}{2\pi^{2}}p_{F}^{2}\frac{dp_{F}}{d\mu} & = & \left[\frac{2\pi^{2}}{d_{g}\mu p_{F}}+\frac{2}{d_{g}}\left(B_{1}-B_{2}+\left(\frac{1}{3}B_{1}+B_{2}\right)\frac{p_{F}^{2}}{\mu^{2}}\right)\right]^{-1}.\label{eq:dndmu1}\end{eqnarray}
If we make the assumption $B_{1,2}\mu^{2}/\pi^{2}\sim B_{1,2}p_{F}^{2}/\pi^{2}\sim\kappa\ll1$
(actually it is a good approximation for realistic values of $B_{1}$,
but not so good for $B_{2}$) together with all other assumptions
in solving Eq. (\ref{eq:dndmu2}), we get an analytical solution for
$\kappa$, \begin{eqnarray}
\kappa(\mu) & = & \left[\kappa(\mu_{0})-\frac{2B_{1}\mu^{2}}{3\pi^{2}}\right]\frac{\mu_{0}^{2}}{\mu^{2}}+\frac{2B_{1}\mu^{2}}{3\pi^{2}},\label{eq:kappa0}\end{eqnarray}
which is in a similar form to Eq. (\ref{eq:kappa-qcd}). The solutions
of $\kappa(\mu)$ also have two branches in analogy to the solutions
of Eq. (\ref{eq:kappa-qcd}) shown in Fig. (\ref{cap:k-mu-qcd}).
We will argue that the lower branch $\kappa(\mu)<\frac{2B_{1}\mu^{2}}{3\pi^{2}}$
is favored, the same situation as in the QCD case. 

If $\kappa$ is small and independent of $\mu$, one can solve Eq.
(\ref{eq:dndmu1}) without the assumption $B_{1,2}\mu^{2}/\pi^{2}\ll1$,
similar to the QCD case, \begin{eqnarray}
\kappa & = & \frac{2B_{1}\mu^{2}}{3\pi^{2}+(7B_{1}+3B_{2})\mu^{2}}.\label{eq:kappa}\end{eqnarray}
 Since we have assumed that $\kappa$ is independent of $\mu$, $B_{1,2}\mu^{2}$
should be constants. The above solution is smaller than the limit
value $\frac{2B_{1}\mu^{2}}{3\pi^{2}}$ of $\kappa(\mu)$ in Eq. (\ref{eq:kappa0})
and recovers it when $B_{1,2}\mu^{2}/\pi^{2}\ll1$. Therefore we have
$\kappa(\mu)<\frac{2B_{1}\mu^{2}}{3\pi^{2}}$ for any values of $\mu$
following Eq. (\ref{eq:kappa0}). Therefore both the QCD and NJL cases
give the consistent trend of $\kappa(\mu)$ as monotonously increasing
functions of $\mu$ towards their limit values at high densities. 

We will use Eq. (\ref{eq:kappa}) in our forthcoming calculations.
The effective mass on the Fermi surface is given by $m\sim\mu\sqrt{2\kappa}$.
In the scalar and pseudoscalar channel, we set $G_{S}=5.1\times10^{-6}\:\mathrm{MeV}^{-2}$,
$\mu=500\:\mathrm{MeV}$ (these parameters are the same as in Ref.
\cite{Ruster:2005jc}), then we obtain $\kappa\approx0.052$ and $m\approx160\:\mathrm{MeV}$.
In the vector channel for $G_{V}=4.5\times10^{-5}\:\mathrm{MeV}^{-2}$,
one obtains $\kappa\approx0.13$ and $m\approx250\:\mathrm{MeV}$.
In both channels, the effective quark masses are much larger than
the current mass but much smaller than $\mu$. 

From Eq. (\ref{eq:kappa-qcd}) or (\ref{eq:kappa}), one also obtains
the angles between $u$ quark and electron momenta and between those
of $u$ and $d$ quarks on the Fermi surface, \begin{equation}
\cos\theta_{ue}\approx1-\kappa,\;\cos\theta_{ud}\approx1-\kappa\frac{\mu_{e}^{2}}{p_{Fu}p_{Fd}}.\label{eq:angles}\end{equation}
These formula are useful in deriving neutrino emissivities in DU processes. 

A few remarks are in order. All the results in this section apply
to DU processes in normal and color superconducting quark matter.
When calculating the quark-quark scattering amplitude, we do not take
into account the fact that the chemical potentials for $u$ and $d$
quarks are different, the same approximation is also made in deriving
Eq. (\ref{eq:pf-mu-qcd}) in QCD \cite{Baym:1975va,Schafer:2004jp}.
In deriving Eq. (\ref{eq:pf-mu}), (\ref{eq:kappa-qcd}) and (\ref{eq:kappa}),
we have made the approximation that $p_{F}\sim\mu$. This is consistent
to the assumption that the effective quark mass is much smaller than
the chemical potential. In this paper we take this assumption and
use Eq. (\ref{eq:kappa}) in forthcoming calculations of neutrino
emissivities. We find in the weak coupling regime the solutions for
the Fermi momentum reduction coefficient $\kappa$ as monotonously
increasing functions of chemical potential in the QCD and NJL cases.
Such a trend in chemical potential implies that the phase space for
neutrino emissions is quenched at lower baryon densities. This property
seems robust and independent of specific models used in computing
the Landau coefficients. In general case where $p_{F}\sim\mu$ does
not hold, Eq. (\ref{eq:pf-mu}) is still valid but $\kappa$ has to
be solved self-consistently. All the results in this section are for
light quarks. If one takes into account strange quarks, the situation
becomes much more complicated and one has to reformulate the whole
Fermi liquid theory. This amounts to dealing with not only scatterings
among strange quarks, but also those between light and strange quarks.
A new Fermi liquid theory with two different Fermi surfaces is necessary.

\section{Quark propagators in a spin-one color superconductor}

To obtain the neutrino emissivity in a spin-one color superconductor,
we calculate the imaginary part of the $W$ boson polarization tensor
from the superconducting quark loop, which involves the quark propagator
$S(K)$. To obtain $S(K)$, we start from $S^{-1}(K)$ as follows
\begin{eqnarray}
S^{-1}(K) & = & \left(\begin{array}{cc}
S_{11}^{-1} & S_{12}^{-1}\\
S_{21}^{-1} & S_{22}^{-1}\end{array}\right)=\left(\begin{array}{cc}
P_{\mu}\gamma^{\mu}+\mu_{f}\gamma_{0}-m & \phi_{f}\gamma_{0}\mathcal{M}_{\mathbf{k}}^{\dagger}\gamma_{0}\\
\phi_{f}\mathcal{M}_{\mathbf{k}} & P_{\mu}\gamma^{\mu}-\mu_{f}\gamma_{0}-m\end{array}\right),\label{eq:s-inverse}\end{eqnarray}
where $\mathcal{M}_{\mathbf{k}}=\Delta_{il}\bm{\gamma}_{\bot}^{i}(\mathbf{k})J_{l}$
and $\gamma_{0}\mathcal{M}_{\mathbf{k}}^{\dagger}\gamma_{0}=\Delta_{il}^{*}\bm{\gamma}_{\bot}^{i}(\mathbf{k})J_{l}$
are $12\times12$ gap matrices in color and Dirac space. $\mu_{f}$
is the chemical potential for quarks of the flavor $f$. $\phi_{f}$
is the magnitude of the gap. The form of $\Delta_{il}$ defines a
specific spin-one phase, see Table I of Ref. \cite{Schmitt:2005wg}
for $\Delta_{il}$ in various phases. Here $\bm{\gamma}_{\bot}(\mathbf{k})\equiv\bm{\gamma}-\widehat{\mathbf{k}}(\widehat{\mathbf{k}}\cdot\bm{\gamma})$
are transverse Dirac matrices perpendicular to a momentum direction
$\widehat{\mathbf{k}}$, and $J_{l}$ are generators of $SO(3)$ algebra
with elements $(J_{l})_{ij}=i\epsilon^{ilj}$ with the anti-symmetric
tensor $\epsilon^{ilj}$. As in Ref. \cite{Schmitt:2005wg}, we only
study transverse phases. 

The structure of $\mathcal{M}_{\mathbf{k}}$ in Eq. (\ref{eq:s-inverse})
looks like in the massless case, but it is different actually. In
the massless case $S_{21}^{-1}$ is expanded in terms of massless
energy projectors $\Lambda_{\mathbf{k},m=0}^{e}$ {[}put $m=0$ in
Eq. (\ref{eq:en-proj}){]} as $S_{21}^{-1}=\Lambda_{\mathbf{k},m=0}^{e}\phi_{f}^{e}\mathcal{M}_{\mathbf{k}}$.
The expansion is justified by the fact that $\Lambda_{\mathbf{k},m=0}^{e}$
is commutable with $\gamma_{5}$ and $\bm{\gamma}_{\bot}^{i}(\mathbf{k})$.
In the massive case the energy projectors $\Lambda_{\mathbf{k}}^{e}$
defined in Eq. (\ref{eq:en-proj}) are \emph{not} commutable with
$\gamma_{5}$ and $\bm{\gamma}_{\bot}(\mathbf{k})$. So if one still
assumes the same form of the gap matrix, the propagators and the dispersion
relations would be much involved as are manifested in Ref. \cite{Aguilera:2005tg,Aguilera:2006cj}. 

The propagator $S(K)$ is found by taking the inverse of $S^{-1}(K)$
in Eq. (\ref{eq:s-inverse}), see Appendix \ref{sec:Energy-projectors}
for the derivation. One then obtains a very transparent form of the
dispersion relations, \begin{equation}
\epsilon_{\mathbf{k},r,f}^{e}=\sqrt{(\mu_{f}-eE_{k})^{2}+\lambda_{\mathbf{k},r}\phi_{f}^{2}},\label{eq:dispersion-rel1}\end{equation}
where $E_{k}=\sqrt{k^{2}+m^{2}}$. $\lambda_{\mathbf{k},r}$ denote
eigenvalues of the matrix $\gamma_{0}\mathcal{M}_{\mathbf{k}}^{\dagger}\mathcal{M}_{\mathbf{k}}\gamma_{0}$
given in Table I of Ref. \cite{Schmitt:2005wg} for various phases.
The gaps in a specific phase are given by $\sqrt{\lambda_{\mathbf{k},r}}\phi_{f}$.
This form is the same as in the massless case except that $k$ is
replaced by $E_{k}$. The results for $S_{11}$ and $S_{22}$ are
given in Eq. (\ref{eq:s11}) and (\ref{eq:s22}). Since only quasi-particles
are relevant in neutrino processes, we only keep excitations of positive
energy, the branch with $e=+$, in $S_{11}$ and $S_{22}$. Then we
have \begin{eqnarray*}
S_{11/22} & = & \frac{k_{0}\mp\mu_{f}\pm E_{k}}{k_{0}^{2}-\epsilon_{\mathbf{k},r,f}^{2}}\mathcal{P}_{\mathbf{k},r}^{\pm}\Lambda_{\mathbf{k}}^{\pm}\gamma_{0},\end{eqnarray*}
where we have suppressed the superscript $e=+$ in the quasi-particle
energy, $\epsilon_{\mathbf{k},r,f}\equiv\epsilon_{\mathbf{k},r,f}^{+}$.
The projectors $\mathcal{P}_{\mathbf{k},r}^{\pm}$ are given in Eq.
(\ref{eq:proj}).

\section{Time derivative of the neutrino distribution function}

Our starting point to compute the neutrino emissivity is the time
derivative of the neutrino distribution function derived from the
Kadanoff-Baym equation for DU processes $u+e^{-}\rightarrow d+\nu$
and $d\rightarrow u+e^{-}+\overline{\nu}$. The procedure can be found
in subsection A of section II in Ref. \cite{Schmitt:2005wg}. The
derivation of Kadanoff-Baym equation can also be found in various
papers using closed time path formalism \cite{Sedrakian:1999jh,Wang:2001dm,Wang:2002qe}.
The result is \begin{eqnarray}
\frac{\partial}{\partial t}f_{\nu}(t,\mathbf{p}_{\nu}) & = & \frac{G_{F}^{2}}{4}\int\frac{d^{3}\mathbf{p}_{e}}{(2\pi)^{3}p_{\nu}p_{e}}L_{\lambda\sigma}(\mathbf{p}_{e},\mathbf{p}_{\nu})n_{F}(p_{e}-\mu_{e})n_{B}(p_{\nu}+\mu_{e}-p_{e})\mathrm{Im}\Pi_{R}^{\lambda\sigma}(Q).\label{eq:dfdt}\end{eqnarray}
Here $n_{F/B}(x)\equiv1/(e^{x}\pm1)$ are Fermi or Bose distribution
functions. $f_{\nu}(t,\mathbf{p}_{\nu})$ is the neutrino (including
the anti-neutrino contribution) distribution function. $\mathbf{p}_{e},\mathbf{p}_{\nu}$
are electron and neutrino momenta respectively. $\mu_{e}$ is the
chemical potential for electrons. $G_{F}$ is the Fermi coupling constant.
The leptonic tensor $L_{\lambda\sigma}(\mathbf{p}_{e},\mathbf{p}_{\nu})$
is given by \begin{eqnarray}
L_{\lambda\sigma}(\mathbf{p}_{e},\mathbf{p}_{\nu}) & = & \mathrm{Tr}\left[\gamma_{\lambda}(1-\gamma^{5})\gamma_{\mu}P_{e}^{\mu}\gamma_{\sigma}(1-\gamma^{5})\gamma_{\rho}P_{\nu}^{\rho}\right]\nonumber \\
 & = & 8\left(i\epsilon_{\lambda\sigma P_{e}P_{\nu}}+P_{\lambda}^{\nu}P_{\sigma}^{e}+P_{\sigma}^{\nu}P_{\lambda}^{e}-g_{\lambda\sigma}P_{e}\cdot P_{\nu}\right),\label{eq:lep-tensor}\end{eqnarray}
where $P_{e}$ and $P_{\nu}$ are on-shell 4-momenta. We have used
the notation $\epsilon_{\lambda\sigma P_{e}P_{\nu}}\equiv\epsilon_{\lambda\sigma\eta\tau}P_{e}^{\eta}P_{\nu}^{\tau}$
with $\epsilon_{\lambda\sigma\eta\tau}$ the anti-symmetric tensor.
$\Pi_{R}^{\lambda\sigma}(Q)$ is the retarded self-energy for $W$
bosons with a superconducting quark loop, where $Q=(p_{e}-\mu_{e}-p_{\nu},\mathbf{p}_{e}-\mathbf{p}_{\nu})$.

\section{Imaginary part of of the W-boson polarization tensor}

Having quark propagators, we can calculate the imaginary part of the
W-boson polarization tensor. Following the same procedure as in Ref.
\cite{Schmitt:2005wg}, one arrives at the following expression, \begin{eqnarray}
\mathrm{Im}\Pi_{R}^{\lambda\sigma}(Q) & = & \frac{T}{V}\mathrm{Im}\sum_{K}\sum_{r,s}\frac{1}{4E_{k}E_{p}}\left\{ \frac{[k_{0}-\mu_{u}+E_{k}][p_{0}-\mu_{d}+E_{p}]}{(k_{0}^{2}-\epsilon_{\mathbf{k},r,u}^{2})(p_{0}^{2}-\epsilon_{\mathbf{p},s,d}^{2})}\mathcal{T}_{rs,+}^{\lambda\sigma}(\widehat{\mathbf{k}},\widehat{\mathbf{p}})\right.\nonumber \\
 &  & \left.+\frac{[k_{0}+\mu_{d}-E_{k}][p_{0}+\mu_{u}-E_{p}]}{(k_{0}^{2}-\epsilon_{\mathbf{k},r,d}^{2})(p_{0}^{2}-\epsilon_{\mathbf{p},s,u}^{2})}\mathcal{T}_{rs,-}^{\lambda\sigma}(\widehat{\mathbf{k}},\widehat{\mathbf{p}})\right\} .\label{eq:polar}\end{eqnarray}
Here we used $P=K+Q$. The quark tensor $\mathcal{T}_{rs,\pm}^{\lambda\sigma}$
involves color and Dirac traces \begin{eqnarray}
\mathcal{T}_{rs,\pm}^{\lambda\sigma}(\widehat{\mathbf{k}},\widehat{\mathbf{p}}) & \equiv & 4E_{k}E_{p}\mathrm{Tr}\left[\gamma^{\lambda}(1\mp\gamma_{5})\mathcal{P}_{\mathbf{k},r}^{\pm}\Lambda_{\mathbf{k}}^{\pm}\gamma^{0}\gamma^{\sigma}(1\mp\gamma_{5})\mathcal{P}_{\mathbf{p},s}^{\pm}\Lambda_{\mathbf{p}}^{\pm}\gamma^{0}\right].\label{eq:t+-}\end{eqnarray}
The negative component $\mathcal{T}_{rs,-}^{\lambda\sigma}$ can be
related to the positive one $\mathcal{T}_{rs,+}^{\lambda\sigma}$
through Eq. (\ref{eq:t-identity}). We can rewrite $\mathcal{T}_{rs,+}^{\lambda\sigma}(\widehat{\mathbf{k}},\widehat{\mathbf{p}})$
as \begin{eqnarray}
\mathcal{T}_{rs,+}^{\lambda\sigma}(\widehat{\mathbf{k}},\widehat{\mathbf{p}}) & = & \mathrm{Tr}\left[\gamma^{\lambda}(1-\gamma_{5})\mathcal{P}_{\mathbf{k},r}^{+}(\gamma_{\mu}K^{\mu})\gamma^{\sigma}(1-\gamma_{5})\mathcal{P}_{\mathbf{p},s}^{+}(\gamma_{\nu}P^{\nu})\right],\label{eq:t+}\end{eqnarray}
where we have dropped all mass terms which are vanishing. One sees
that $\gamma_{\mu}K^{\mu}$ and $\gamma_{\nu}P^{\nu}$ appear in $\mathcal{T}_{rs,+}^{\lambda\sigma}(\widehat{\mathbf{k}},\widehat{\mathbf{p}})$,
where $K^{\mu}$ and $P^{\nu}$ are on-shell quark momenta. In normal
quark matter, one of the projectors, say $\mathcal{P}_{\mathbf{k},r=1}^{+}$
can be set to 1 and others with $r\neq1$ set to zero, then one reproduces
the result in the normal phase, \begin{eqnarray}
\mathcal{T}^{\lambda\sigma}(K,P) & \equiv & \mathrm{Tr}\left[\gamma^{\lambda}(1-\gamma_{5})(\gamma_{\mu}K^{\mu})\gamma^{\sigma}(1-\gamma_{5})(\gamma_{\nu}P^{\nu})\right]\nonumber \\
 & = & 8\left(i\epsilon^{\lambda\sigma KP}+P^{\lambda}K^{\sigma}+P^{\sigma}K^{\lambda}-g^{\lambda\sigma}K\cdot P\right).\label{eq:quark-tensor-nm}\end{eqnarray}
Note that the anomalous propagators $S_{21}$ and $S_{12}$ containing
the $uu$ and $dd$ condensates do not contribute to the self-energy.
This is in contrast to spin-zero color superconductors, such as the
2SC and CFL phases, in which the anomalous propagators are off-diagonal
in flavor space. This difference is related to the conservation of
electric charge. Performing the Mastubara sum, one can extract the
imaginary part, \begin{eqnarray}
\mathrm{Im}\Pi_{R}^{\lambda\sigma}(Q) & = & -\pi\sum_{r,s}\sum_{e_{1},e_{2}}\int\frac{d^{3}\mathbf{k}}{(2\pi)^{3}}\frac{1}{4E_{k}E_{p}}\nonumber \\
 &  & \times\left[\mathcal{T}_{rs,+}^{\lambda\sigma}(\widehat{\mathbf{k}},\widehat{\mathbf{p}})B_{\mathbf{k},r,u}^{e_{1}}B_{\mathbf{p},s,d}^{e_{2}}\frac{n_{F}(-e_{1}\epsilon_{\mathbf{k},r,u})n_{F}(e_{2}\epsilon_{\mathbf{p},s,d})}{n_{B}(-e_{1}\epsilon_{\mathbf{k},r,u}+e_{2}\epsilon_{\mathbf{p},s,d})}\delta(q_{0}-e_{1}\epsilon_{\mathbf{k},r,u}+e_{2}\epsilon_{\mathbf{p},s,d})\right.\nonumber \\
 &  & \left.+\mathcal{T}_{rs,-}^{\lambda\sigma}(\widehat{\mathbf{k}},\widehat{\mathbf{p}})B_{\mathbf{k},r,d}^{e_{1}}B_{\mathbf{p},s,u}^{e_{2}}\frac{n_{F}(e_{1}\epsilon_{\mathbf{k},r,d})n_{F}(-e_{2}\epsilon_{\mathbf{p},s,u})}{n_{B}(e_{1}\epsilon_{\mathbf{k},r,d}-e_{2}\epsilon_{\mathbf{p},s,u})}\delta(q_{0}+e_{1}\epsilon_{\mathbf{k},r,d}-e_{2}\epsilon_{\mathbf{p},s,u})\right],\label{eq:im1}\end{eqnarray}
where $q_{0}=p_{e}-\mu_{e}-p_{\nu}$. The Bogoliubov coefficients
are defined by \begin{eqnarray*}
B_{\mathbf{k},r,f}^{e} & = & \frac{\epsilon_{\mathbf{k},r,f}+e(\mu_{f}-E_{k})}{2\epsilon_{\mathbf{k},r,f}},\end{eqnarray*}
with $f=u,d$ and $e=\pm$. The two terms inside the square brackets
are charge-conjugate of each other and give the same result. To see
this, one changes the summation indices $e_{1}\leftrightarrow e_{2},$$r\leftrightarrow s$
in the second term, introduces the new integral variable $\mathbf{k}\rightarrow-\mathbf{k}-\mathbf{q}=-\mathbf{p}$
(also means $\mathbf{p}=\mathbf{k}+\mathbf{q}\rightarrow-\mathbf{k}$),
and uses \begin{eqnarray}
\mathcal{T}_{sr,-}^{\lambda\sigma}(-\widehat{\mathbf{p}},-\widehat{\mathbf{k}}) & = & \mathcal{T}_{rs,+}^{\lambda\sigma}(\widehat{\mathbf{k}},\widehat{\mathbf{p}}).\label{eq:t-identity}\end{eqnarray}
The proof of the above identity for all phases considered in this
paper is given in Appendix \ref{sec:proof}. Taking into account $\lambda_{\mathbf{k},r}=\lambda_{-\mathbf{k},r}$
and then $\epsilon_{\mathbf{k},r,f}=\epsilon_{-\mathbf{k},r,f}$,
we end up with the first term in Eq. (\ref{eq:im1}). So we can keep
the first term in Eq. (\ref{eq:im1}) and double the result, \begin{eqnarray}
\mathrm{Im}\Pi_{R}^{\lambda\sigma}(Q) & = & -2\pi\sum_{r,s}\sum_{e_{1},e_{2}}\int\frac{d^{3}\mathbf{k}}{(2\pi)^{3}}\frac{1}{4E_{k}E_{p}}\mathcal{T}_{rs,+}^{\lambda\sigma}(\widehat{\mathbf{k}},\widehat{\mathbf{p}})B_{\mathbf{k},r,u}^{e_{1}}B_{\mathbf{p},s,d}^{e_{2}}\nonumber \\
 &  & \times\frac{n_{F}(-e_{1}\epsilon_{\mathbf{k},r,u})n_{F}(e_{2}\epsilon_{\mathbf{p},s,d})}{n_{B}(-e_{1}\epsilon_{\mathbf{k},r,u}+e_{2}\epsilon_{\mathbf{p},s,d})}\delta(q_{0}-e_{1}\epsilon_{\mathbf{k},r,u}+e_{2}\epsilon_{\mathbf{p},s,d}).\label{eq:im2}\end{eqnarray}

\section{Quark mass corrections to Neutrino emissivity}

In this section we will calculate neutrino emissivities with massive
quarks and identify the mass corrections. We will use the effective
quark mass $m\sim\mu\sqrt{2\kappa}$ given in Sect. \ref{sec:phase-space}.
The effective quark mass is formed on the Fermi surface by the interquark
interaction in the Landau Fermi liquid. Normally the effective quark
mass is much larger than the current mass, but it is still small compared
to the large chemical potential. 

Inserting $\mathrm{Im}\Pi_{R}^{\lambda\sigma}(Q)$ in Eq. (\ref{eq:im2})
back into Eq. (\ref{eq:dfdt}), one obtains \begin{eqnarray}
\frac{\partial}{\partial t}f_{\nu}(t,\mathbf{p}_{\nu}) & = & -\frac{\pi}{8}G_{F}^{2}\int\frac{d^{3}\mathbf{p}}{(2\pi)^{3}p_{\nu}p_{e}}\int\frac{d^{3}\mathbf{k}}{(2\pi)^{3}E_{k}E_{p}}\sum_{r,s}\sum_{e_{1},e_{2}}L_{\lambda\sigma}(\mathbf{p}_{e},\mathbf{p}_{\nu})\mathcal{T}_{rs,+}^{\lambda\sigma}(\widehat{\mathbf{k}},\widehat{\mathbf{p}})B_{\mathbf{k},r,u}^{e_{1}}B_{\mathbf{p},s,d}^{e_{2}}\nonumber \\
 &  & \times n_{F}(p_{e}-\mu_{e})n_{F}(-e_{1}\epsilon_{\mathbf{k},r,u})n_{F}(e_{2}\epsilon_{\mathbf{p},s,d})\delta(q_{0}-e_{1}\epsilon_{\mathbf{k},r,u}+e_{2}\epsilon_{\mathbf{p},s,d}),\label{eq:dfdt1}\end{eqnarray}
where we have replaced the integral over $\mathbf{p}_{e}$ with that
over $\mathbf{p}$. In normal quark matter, the quark tensor $\mathcal{T}^{\lambda\sigma}$
in Eq. (\ref{eq:quark-tensor-nm}) contracts with the leptonic tensor
$L_{\lambda\sigma}$ in Eq. (\ref{eq:lep-tensor}) giving the matrix
element of the DU processes \begin{eqnarray}
\mathcal{T}^{\lambda\sigma}L_{\lambda\sigma} & = & |M|^{2},\label{eq:tr-normal}\end{eqnarray}
with $|M|^{2}=256(P_{e}\cdot K)(P_{\nu}\cdot P)$. Using Eq. (\ref{eq:pf-mu})
and (\ref{eq:angles}), the matrix element $|M|^{2}$ can be worked
out, \begin{eqnarray}
|M|^{2} & \approx & 512\kappa E_{k}E_{p}p_{e}p_{\nu}\left[1-(1-\kappa)\cos\theta_{d\nu}\right],\end{eqnarray}
with $\kappa$ given in Eq. (\ref{eq:kappa}). 

The quark tensor $\mathcal{T}_{rs,+}^{\lambda\sigma}(\widehat{\mathbf{k}},\widehat{\mathbf{p}})$
in spin-one phases are evaluated in Appendix \ref{sec:quark-tensor}.
Using Eq. (\ref{eq:mass-t}), its contraction with $L_{\lambda\sigma}$
is given by \begin{eqnarray}
L_{\lambda\sigma}\mathcal{T}_{rs,+}^{\lambda\sigma} & = & \omega_{rs}|M|^{2}+L_{\lambda\sigma}\delta\mathcal{T}_{rs}^{\lambda\sigma}.\label{eq:matrix-el-mass}\end{eqnarray}
Here the values of $\omega_{rs}$ are the same as in the massless
case (see Tab. II of Ref. \cite{Schmitt:2005wg} or Appendix \ref{sec:quark-tensor}).
As is shown in Appendix \ref{sec:quark-tensor}, the dominant contribution
in $\delta\mathcal{T}_{rs}^{\lambda\sigma}$ is from the term proportional
to $R_{k}\mathcal{T}^{\lambda\sigma}(\widetilde{K}_{0},P)$ with $R_{k}\equiv(E_{k}-k)/k\approx m^{2}/(2p_{Fu}^{2})\sim\kappa$
and $\widetilde{K}_{0}\equiv(k,-\mathbf{k})$, because $P_{e}\cdot\widetilde{K}_{0}\sim2p_{e}k$
is much larger than $P_{e}\cdot K\sim2\kappa E_{k}p_{e}$. So $\delta\mathcal{T}_{rs}^{\lambda\sigma}$
can be written as $\delta\mathcal{T}_{rs}^{\lambda\sigma}\approx\chi_{rs}R_{k}\mathcal{T}^{\lambda\sigma}(\widetilde{K}_{0},P)$
where $\chi_{rs}$ is listed in Tab. \ref{cap:tab1}. Terms of $R_{p}\mathcal{T}^{\lambda\sigma}(K,\widetilde{P}_{0})\sim\kappa m^{2}/p_{Fd}^{2}\sim\kappa^{2}$
are next to leading order since the appearance of $\widetilde{P}_{0}$
does not enhance the order of magnitude of the result, so they are
suppressed by $\kappa$ compared to the massless term. Also terms
of $R_{k}R_{p}\mathcal{T}^{\lambda\sigma}(\widetilde{K}_{0},\widetilde{P}_{0})\sim m^{4}/(p_{Fu}^{2}p_{Fd}^{2})\sim\kappa^{2}$
are also suppressed due to double $R$ factors. Explicitly one derives
\begin{eqnarray}
L_{\lambda\sigma}\delta\mathcal{T}_{rs}^{\lambda\sigma} & \approx & \chi_{rs}\frac{m^{2}}{2p_{Fu}\mu_{u}}256E_{k}E_{p}p_{e}p_{\nu}\left[1-(1-\kappa)\cos\theta_{d\nu}\right],\nonumber \\
 & \approx & \chi_{rs}\frac{\kappa}{1-\kappa}256E_{k}E_{p}p_{e}p_{\nu}\left[1-(1-\kappa)\cos\theta_{d\nu}\right].\label{eq:nm-t}\end{eqnarray}
The term $\kappa\cos\theta_{d\nu}$ inside the square brackets can
be dropped since it is of high order. 

Let us understand Tab. \ref{cap:tab1}. We find $\chi_{rs}=0$ for
the polar phase. The reason is simple: all projectors (\ref{eq:proj-polar})
are in color space or decoupled from Dirac space, so the quark tensor
$\mathcal{T}_{rs,+}^{\lambda\sigma}$ is proportional to $\mathcal{T}^{\lambda\sigma}$.
For other phases, A, planar and CSL, color and Dirac indices are coupled.
Thus with massive quarks, the Dirac structure of the quark tensor
is more complicated. Taking the A phase for example, it is easy to
understand $\chi_{rs}=0$ for $rs=13,31,23,32,33$, because the excitation
branch 3 only involves the color \emph{blue} while branch 1 and 2
involve colors \emph{red} and \emph{green}. One observes that $\chi_{12}=\chi_{21}\neq0$
even in the collinear limit, meaning that there is an entanglement
between two different branches. Remembering that the projectors of
the A phase can be expressed in terms of helicity projectors, as shown
in Eq. (60) of Ref. \cite{Schmitt:2005wg}, quasiparticles of branch
1 have helicity +1 for $\widehat{k}_{3}<0$ and helicity -1 for $\widehat{k}_{3}>0$,
while quasiparticles of branch 2 have opposite helicities. In the
massless case, helicity means chirality. The DU processes only allow
quarks with left chirality to participate. Therefore in the collinear
limit $\widehat{k}_{3}\approx\widehat{p}_{3}$, different branches
cannot crosstalk. However when quarks have masses, a helicity eigenstate
is not chirality one. In an eigenstate of left chirality, one can
find eigenstates of both helicities, which makes two different branches
coupled together. In the same way, one can understand the appearance
$\delta\mathcal{T}_{rs}^{\lambda\sigma}$ in the quark tensor with
different structure from $\mathcal{T}^{\lambda\sigma}$. A projector
of a spin-one phase can be expanded with respect to helicity projectors.
In the massless case, the quark tensors for all branches have the
same Dirac structure as in normal quark matter. With non-zero quark
mass, this is no longer the case: new terms in the quark tensor emerge
from an overlapping of subspace with different helicities (but with
the same chirality). They have different structure and cannot be collected
into $\mathcal{T}^{\lambda\sigma}$ as in the massless case. This
can be seen from the fact that such terms as $\gamma_{0}\bm{\gamma}\cdot\widehat{\mathbf{k}}\gamma_{\mu}K^{\mu}$
or $\gamma_{\mu}K^{\mu}\bm{\gamma}\cdot\widehat{\mathbf{k}}\gamma_{0}$
appear inside the trace of Eq. (\ref{eq:t+}) for A, planar and CSL
phases, \begin{eqnarray}
\gamma_{0}\bm{\gamma}\cdot\widehat{\mathbf{k}}\gamma_{\mu}K^{\mu} & = & \gamma_{\mu}K^{\mu}(\bm{\gamma}\cdot\widehat{\mathbf{k}})\gamma_{0}=\gamma_{\mu}K^{\mu}-R_{k}\gamma_{\mu}\widetilde{K}_{0}^{\mu}.\end{eqnarray}
When masses are zero, the result is just $\gamma_{\mu}K^{\mu}$, which
results in traces proportional to $\mathcal{T}^{\lambda\sigma}$ in
Eq. (\ref{eq:quark-tensor-nm}). When masses are non-zero, there is
an additional term $R_{k}\gamma_{\mu}\widetilde{K}_{0}^{\mu}$ different
from $\mathcal{T}^{\lambda\sigma}$. 

\begin{table}

\caption{\label{cap:tab1}The values of $\chi_{rs}$ in quark mass term $\delta\mathcal{T}_{rs}^{\lambda\sigma}\approx\chi_{rs}R_{k}\mathcal{T}^{\lambda\sigma}(\widetilde{K}_{0},P)$.
For the polar phase $\chi_{rs}=0$ for all $r$ and $s$. The row
and column are labeled by $r$ and $s$ respectively. The collinear
limit is taken. We used the notation $\xi\equiv\widehat{k}_{3}$ and
$\theta(\xi)\equiv1+\mathrm{sgn}(\xi)$ for the step function of $\xi$. }

\begin{tabular}{|c|c|c|}
\hline 
Planar&
1&
2\tabularnewline
\hline 
1&
$-\frac{2\xi^{2}}{(1+\xi^{2})^{2}}$&
$\frac{2\xi^{2}}{(1+\xi^{2})^{2}}$\tabularnewline
\hline 
2&
$\frac{2\xi^{2}}{(1+\xi^{2})^{2}}$&
$-\frac{2\xi^{2}}{(1+\xi^{2})^{2}}$\tabularnewline
\hline
\end{tabular}\hspace{1cm}\begin{tabular}{|c|c|c|c|}
\hline 
A&
1&
2&
3\tabularnewline
\hline 
1&
$-\theta(\xi)$&
$\theta(-\xi)$&
0\tabularnewline
\hline 
2&
$\theta(\xi)$&
$-\theta(-\xi)$&
0\tabularnewline
\hline 
3&
0&
0&
0\tabularnewline
\hline
\end{tabular}\hspace{1cm}\begin{tabular}{|c|c|c|}
\hline 
CSL&
1&
2\tabularnewline
\hline 
1&
$-\frac{1}{2}$&
$\frac{1}{2}$\tabularnewline
\hline 
2&
$\frac{1}{2}$&
$-\frac{1}{2}$\tabularnewline
\hline
\end{tabular}
\end{table}

Using Eq. (\ref{eq:dfdt1}) and (\ref{eq:nm-t}), the emissivity is
written as a sum of two terms \begin{eqnarray}
\epsilon_{\nu} & = & -\int\frac{d^{3}\mathbf{p}_{\nu}}{(2\pi)^{3}}p_{\nu}\frac{\partial}{\partial t}f_{\nu}(t,\mathbf{p}_{\nu})\equiv\epsilon_{\nu}^{(0)}+\epsilon_{\nu}^{(1)}.\label{eq:em}\end{eqnarray}
Here $\epsilon_{\nu}^{(0)}$ is in the same form as the emissivity
in the massless case which we denote as $\epsilon_{\nu}^{(m=0)}$,
i.e. Eq. (40) of Ref. \cite{Schmitt:2005wg}, except that all $k$
and $p$ are replaced by $E_{k}$ and $E_{p}$ in dispersion relations
respectively, and that $2\alpha_{S}/(3\pi)$ replaced by $\kappa$
in Eq. (\ref{eq:kappa-qcd}) or (\ref{eq:kappa}). It comes from the
term $\omega_{rs}|M|^{2}$ in Eq. (\ref{eq:matrix-el-mass}). In the
massless case, the ranges of the integrals over $k-\mu_{u}$ and/or
$p-\mu_{d}$ on which the quark distributions depend are $[-\mu_{u/d},\infty]$,
which are further taken to be $[-\infty,\infty]$. With non-zero quark
mass, the integrals over $k-\mu_{u}$ and/or $p-\mu_{d}$ have to
be transformed to those over $E_{k/p}-\mu_{u/d}$ on which the quark
distribution functions depend, i.e. \begin{eqnarray}
dpdk & = & (E_{p}E_{k}/pk)dE_{p}dE_{k}=(E_{k}E_{p}/pk)d(E_{p}-\mu)d(E_{k}-\mu)\nonumber \\
 & \approx & \frac{\mu_{u}\mu_{d}}{pk}d(E_{p}-\mu)d(E_{k}-\mu).\label{eq:dkdp}\end{eqnarray}
In the last equality we have chosen $E_{k}\approx\mu_{u}$ and $E_{p}\approx\mu_{d}$
on the Fermi surface. The range of $E_{k/p}-\mu_{u/d}$ becomes $[m-\mu_{u/d},\infty]$
and is further approximated to be $[-\infty,\infty]$ if $m\ll\mu_{u/d}$.
Using Eq. (\ref{eq:angles}), (\ref{eq:dfdt1}) and (\ref{eq:matrix-el-mass}),
we obtain \begin{equation}
\epsilon_{\nu}^{(0)}=\epsilon_{\nu}^{(m=0)}.\label{eq:mass0}\end{equation}
This means $\epsilon_{\nu}^{(0)}$ is identical to the emissivity
in the massless case. 

\begin{figure}

\caption{\label{cap:g-vs-phi}$G'$ versus $\varphi$ for spin-one phases. }

\includegraphics[%
  scale=0.9]{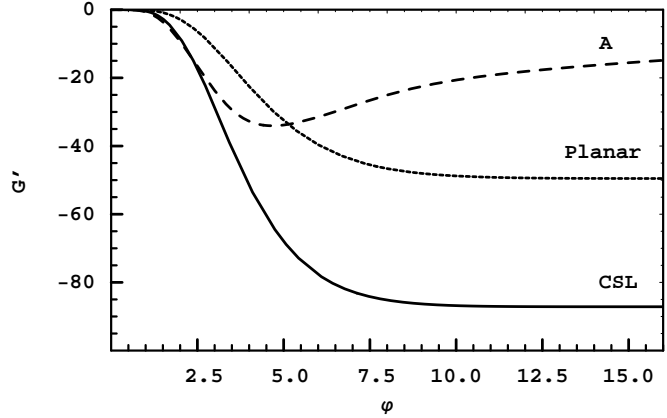}
\end{figure}

The second part of the emissivity in Eq. (\ref{eq:em}), $\epsilon_{\nu}^{(1)}$,
has obvious mass dependence. It comes from the term $L_{\lambda\sigma}\delta\mathcal{T}_{rs}^{\lambda\sigma}$
in Eq. (\ref{eq:matrix-el-mass}). Inserting Eq. (\ref{eq:nm-t})
into Eq. (\ref{eq:dfdt1}) and using Eq. (\ref{eq:em}), one obtains
\begin{eqnarray}
\epsilon_{\nu}^{(1)} & = & 32\pi\frac{\kappa}{1-\kappa}G_{F}^{2}\mu_{e}\mu_{u}\mu_{d}\int\frac{d^{3}\mathbf{p}_{\nu}}{(2\pi)^{3}}p_{\nu}\int\frac{dE_{p}d\Omega_{\mathbf{p}}}{(2\pi)^{3}}\int\frac{dE_{k}d\Omega_{\mathbf{k}}}{(2\pi)^{3}}\sum_{r,s}\sum_{e_{1},e_{2}}\chi_{rs}(1-\cos\theta_{d\nu})\nonumber \\
 &  & \times B_{\mathbf{k},r,u}^{e_{1}}B_{\mathbf{p},s,d}^{e_{2}}n_{F}(p_{e}-\mu_{e})n_{F}(-e_{1}\epsilon_{\mathbf{k},r,u})n_{F}(e_{2}\epsilon_{\mathbf{p},s,d})\delta[\cos\theta_{ud}-1+\kappa\mu_{e}^{2}/(p_{Fu}p_{Fd})],\end{eqnarray}
where Eq. (\ref{eq:dkdp}) was used in the integral. We have also
used \begin{eqnarray}
\delta(p_{e}-\mu_{e}-p_{\nu}-e_{1}\epsilon_{\mathbf{k},r,u}+e_{2}\epsilon_{\mathbf{p},s,d}) & \approx & \frac{\mu_{e}}{p_{Fu}p_{Fd}}\delta[\cos\theta_{ud}-1+\kappa\mu_{e}^{2}/(p_{Fu}p_{Fd})].\end{eqnarray}
After taking the same procedure as in Ref. \cite{Schmitt:2005wg},
one has \begin{eqnarray}
\epsilon_{\nu}^{(1)} & \approx & \frac{1}{\pi^{5}}\frac{\kappa}{1-\kappa}G_{F}^{2}\mu_{e}\mu_{u}\mu_{d}T^{6}G'(\varphi_{u},\varphi_{d}),\label{eq:mass-corr}\end{eqnarray}
with integral $G'$ defined by \begin{eqnarray}
G'(\varphi_{u},\varphi_{d}) & \equiv & \sum_{e_{1},e_{2}}\sum_{r,s}\int_{-1}^{1}d\xi\chi_{rs}(\xi)\int_{0}^{\infty}dvv^{3}\int_{0}^{\infty}dx\int_{0}^{\infty}dy\; n_{F}(v+e_{1}\sqrt{y^{2}+\lambda_{r}(\xi)\varphi_{u}^{2}}-e_{2}\sqrt{x^{2}+\lambda_{s}(\xi)\varphi_{d}^{2}})\nonumber \\
 &  & \times n_{F}(-e_{1}\sqrt{y^{2}+\lambda_{r}(\xi)\varphi_{u}^{2}})n_{F}(e_{2}\sqrt{x^{2}+\lambda_{s}(\xi)\varphi_{d}^{2}}).\label{eq:G'}\end{eqnarray}
Here we used dimensionless gaps scaled by the temperature, $\varphi_{u,d}\equiv\phi_{u,d}/T$.
Eq. (\ref{eq:mass-corr}) shows that $\epsilon_{\nu}^{(1)}$ is in
the same order as $\epsilon_{\nu}^{(m=0)}$. We can take the special
case $\varphi_{u}=\varphi_{d}=\varphi$ and do the numerical calculation
for $G'$ as a function of $\varphi$. The results are shown in Fig.
(\ref{cap:g-vs-phi}). We see that $G'<0$ for all three spin-one
phases. At asymptotically low temperature or large $\varphi$, $G'$
for CSL and Planar phases approach constants due to the gapless mode
2, \[
G'=-\frac{457\pi^{6}}{5040}\;\;(\mathrm{CSL}),\;-\frac{457\pi^{6}(\pi/2-1)}{5040}\;\;(\mathrm{Planar}).\]
For the A phase, $G'$ turns to zero at large $\varphi$ since all
contributions come from the gapped modes. 

In summary, from Eq. (\ref{eq:em}), (\ref{eq:mass0}) and (\ref{eq:mass-corr}),
the emissivity can be written as the sum of two terms, \begin{equation}
\epsilon_{\nu}=\epsilon_{\nu}^{m=0}+\epsilon_{\nu}^{(1)},\label{eq:em-final}\end{equation}
where $\epsilon_{\nu}^{(1)}$ is the correction due to non-zero quark
mass and it is of the same order as $\epsilon_{\nu}^{(m=0)}$. It
is zero for the polar phase. For the CSL and Planar phases, we obtain
\[
\epsilon_{\nu}^{(1)}\approx C_{i}\frac{1}{1-\kappa}\epsilon_{\nu}^{(m=0)}\sim C_{i}\epsilon_{\nu}^{(m=0)},\;\; i=\mathrm{A,\: CSL,\: Planar}\]
 At asymptotically large $\varphi$, the coefficients $C_{i}$ are
given by \begin{eqnarray}
C_{\mathrm{A}} & = & 0,\;\; C_{\mathrm{CSL}}=-\frac{1}{4}\approx-0.25,\;\; C_{\mathrm{Planar}}=-\frac{(\pi-2)}{8}\approx-0.15.\label{eq:numbers}\end{eqnarray}
which show that mass corrections are about 25\% (CSL) and 15\% (Planar)
of the contributions in the massless case at very low temperatures.
These results are quite universal at $\kappa\ll1$, which is the case
for the weak coupling approach in QCD with one gluon exchange as $\kappa\sim\alpha_{S}$.
This is also the case for the realistic values of coupling constants
$G_{S}\sim5.1\times10^{-6}\:\mathrm{MeV}^{-2}$ and $G_{V}\sim4.5\times10^{-5}\:\mathrm{MeV}^{-2}$
in the NJL model which give $\kappa\sim0.05-0.13$. 

A few comments are in order. The gap matrix in Eq. (\ref{eq:s-inverse})
leads to the same form of the dispersion relations as in the massless
case. This much facilitates the calculation with transparency. If
the gap matrix is not in this form, the dispersion relation is more
involved \cite{Aguilera:2005tg,Aguilera:2006cj}. But one can prove
that they are equivalent in some limits (see the discussion in section
VII of Ref. \cite{Schmitt:2005wg}). We have used the collinear limit,
where the momentum direction of $u$ quarks is almost the same as
that of $d$ quarks. The effective quark mass $m\sim\mu\sqrt{2\kappa}$
is also used in the calculation, which is much larger than the current
mass but small compared to the chemical potential. Then we find that
the mass corrections to the emissivity are of the same order as in
the massless case. The mass corrections from gapless modes in CSL
and Planar phases which are dominant at low temperatures are about
25\% and 15\% of the contributions in the massless case respectively.
These percentages for the mass corrections are universal as long as
$\kappa\ll1$. The mass corrections could be very different near the
chiral phase transition and for the DU processes with strange quarks
with even larger quark mass. However the current formulation is only
valid at small mass limit compared to the chemical potential. For
very large quark mass, the Landau Fermi liquid theory has to be reformulated
and the integral in the emissivity is very different from the massless
case since the integral range $E-\mu$ cannot literally be taken as
$[-\infty,\infty]$ due to $m\sim\mu$. Not only this has to be modified,
one also has to deal with scatterings between light and strange quarks,
which involves the Fermi liquid theory with two different Fermi surfaces.
This is beyond the scope of the current work and deserves a future
study. 

\begin{acknowledgments} Q.W. thanks D.~Blaschke, D.~H.~Rischke, A.~Schmitt and I.~A.~Shovkovy for comments and discussions. Q.W. is supported in part by the startup grant from University of Science and Technology of China (USTC) in association with 'Bai Ren' project of Chinese Academy of Sciences (CAS). Z.G.W. is supported by National Natural Science Foundation of China (NSFC) under grant 10405009. 

\end{acknowledgments}  

 \appendix

\section{Energy projectors and propagators }

\label{sec:Energy-projectors}We make use of energy projectors to
get $S(K)$ from $S^{-1}(K)$ in Eq. (\ref{eq:s-inverse}). The energy
projectors are given by \begin{eqnarray}
\Lambda_{\mathbf{k}}^{e} & = & \frac{1}{2}\left[1+e\frac{\gamma_{0}(\bm{\gamma}\cdot\mathbf{k}+m)}{E_{k}}\right],\label{eq:en-proj}\end{eqnarray}
where $E_{k}=\sqrt{k^{2}+m^{2}}$. We can express $S_{11}^{-1}$ and
$S_{22}^{-1}$ in terms of energy projectors: \[
S_{11/22}^{-1}=(k_{0}\pm\mu_{f}-eE_{k})\gamma_{0}\Lambda_{\mathbf{k}}^{e}.\]
The chemical potential $\mu_{f}$ has flavor index $f$. The elements
of $S(K)$ are given by \begin{eqnarray*}
S_{11} & = & \left[S_{11}^{-1}-S_{12}^{-1}(S_{22}^{-1})^{-1}S_{21}^{-1}\right]^{-1},\\
S_{22} & = & \left[S_{22}^{-1}-S_{21}^{-1}(S_{11}^{-1})^{-1}S_{12}^{-1}\right]^{-1},\\
S_{21} & = & -(S_{22}^{-1})^{-1}S_{21}^{-1}S_{11},\\
S_{12} & = & -(S_{11}^{-1})^{-1}S_{12}^{-1}S_{22}.\end{eqnarray*}
To get $S_{11}$, we need to evaluate \begin{eqnarray*}
S_{12}^{-1}(S_{22}^{-1})^{-1}S_{21}^{-1} & = & \phi_{f}^{2}\gamma_{0}\alpha_{\bot}^{i}(\mathbf{k})\Delta_{il}^{*}J_{l}(S_{22}^{-1})^{-1}\gamma_{0}\alpha_{\bot}^{i'}(\mathbf{k})\Delta_{i'l'}J_{l'}\\
 & = & \phi_{f}^{2}\gamma_{0}[\alpha_{\bot}^{i}(\mathbf{k})\Delta_{il}^{*}J_{l}][\alpha_{\bot}^{i'}(\mathbf{k})\Delta_{i'l'}J_{l'}]\frac{1}{k_{0}-\mu_{f}-eE_{k}}\Lambda_{\mathbf{k}}^{-e}.\end{eqnarray*}
Here we have used $\Lambda_{\mathbf{k}}^{e}\alpha_{\bot}^{i}(\mathbf{k})=\alpha_{\bot}^{i}(\mathbf{k})\Lambda_{\mathbf{k}}^{-e}$
with $\bm{\alpha}=\gamma_{0}\bm{\gamma}$. Note that \[
L_{\mathbf{k}}^{+}\equiv[\alpha_{\bot}^{i}(\mathbf{k})\Delta_{il}^{*}J_{l}][\alpha_{\bot}^{i'}(\mathbf{k})\Delta_{i'l'}J_{l'}]\]
 is just the matrix $\gamma_{0}\mathcal{M}_{\mathbf{k}}^{\dagger}\mathcal{M}_{\mathbf{k}}\gamma_{0}$
in Eq. (14) of Ref. \cite{Schmitt:2005wg}, and it is commutable with
$\Lambda_{\mathbf{k}}^{e}$. We can decompose $L_{\mathbf{k}}^{+}$
in terms of its projectors as $L_{\mathbf{k}}^{+}=\lambda_{\mathbf{k},r}\mathcal{P}_{\mathbf{k},r}^{+}$.
Then we obtain \begin{eqnarray}
S_{11} & = & \frac{k_{0}-\mu_{f}+eE_{k}}{k_{0}^{2}-(\epsilon_{\mathbf{k},r,f}^{e})^{2}}\mathcal{P}_{\mathbf{k},r}^{+}\Lambda_{\mathbf{k}}^{e}\gamma_{0},\label{eq:s11}\end{eqnarray}
with the quasi-particle excitation energy $\epsilon_{\mathbf{k},r,f}^{e}=\sqrt{(\mu-eE_{k})^{2}+\lambda_{\mathbf{k},r}\phi_{f}^{2}}$. 

Making the change $\mu_{f}\rightarrow-\mu_{f}$ in $S_{11}$, we get
$S_{22}$ \begin{equation}
S_{22}=\frac{k_{0}+\mu_{f}-eE_{k}}{k_{0}^{2}-(\epsilon_{\mathbf{k},r,f}^{e})^{2}}\mathcal{P}_{\mathbf{k},r}^{-}\Lambda_{\mathbf{k}}^{-e}\gamma_{0},\label{eq:s22}\end{equation}
where $\mathcal{P}_{\mathbf{k},r}^{-}$ are projectors of \[
L_{\mathbf{k}}^{-}\equiv[\alpha_{\bot}^{i}(\mathbf{k})\Delta_{il}J_{l}][\alpha_{\bot}^{i'}(\mathbf{k})\Delta_{i'l'}^{*}J_{l'}],\]
 whose eigenvalues $\lambda_{\mathbf{k},r}$ are the same as $L_{\mathbf{k}}^{+}$'s.
The projectors $\mathcal{P}_{\mathbf{k},r}^{\pm}$ are given by \begin{eqnarray}
\mathcal{P}_{\mathbf{k},t}^{\pm} & = & \prod_{r\neq t}\frac{L_{\mathbf{k}}^{\pm}-\lambda_{\mathbf{k},r}}{\lambda_{\mathbf{k},t}-\lambda_{\mathbf{k},r}}.\label{eq:proj}\end{eqnarray}
Note that the projectors $\mathcal{P}_{\mathbf{k},r}^{\pm}$ are commutable
with $\Lambda_{\mathbf{k}}^{e}$ since $[L_{\mathbf{k}}^{\pm},\Lambda_{\mathbf{k}}^{e}]=0$.

\section{Proof of the identity (\ref{eq:t-identity})}

\label{sec:proof}One starts from $\mathcal{T}_{rs,-}^{\lambda\sigma}$
as follows \[
\mathcal{T}_{sr,-}^{\lambda\sigma}(-\widehat{\mathbf{p}},-\widehat{\mathbf{k}})=\mathrm{Tr}\left[\gamma^{\lambda}(1+\gamma_{5})\mathcal{P}_{-\mathbf{p},s}^{-}\Lambda_{-\mathbf{p}}^{-}\gamma^{0}\gamma^{\sigma}(1+\gamma_{5})\mathcal{P}_{-\mathbf{k},r}^{-}\Lambda_{-\mathbf{k}}^{-}\gamma^{0}\right].\]
Inserting the charge conjugate operator $C=i\gamma_{0}\gamma^{2}$
and using $C\alpha^{i}\alpha^{j}C^{-1}=(\alpha^{j}\alpha^{i})^{T}$,
$\alpha_{\bot}^{i}(-\mathbf{k})=\alpha_{\bot}^{i}(\mathbf{k})$ and
$J_{l}=-J_{l}^{T}$, one obtains \begin{eqnarray*}
C\mathcal{P}_{-\mathbf{k},r}^{-}C^{-1} & = & (\mathcal{P}_{\mathbf{k},r}^{+})^{T}.\end{eqnarray*}
One uses $C\gamma^{\mu}C^{-1}=-\gamma^{\mu T}$ to obtain \begin{eqnarray*}
C\Lambda_{-\mathbf{p}}^{-}\gamma^{0}C^{-1} & = & (\Lambda_{\mathbf{p}}^{+}\gamma^{0})^{T}.\end{eqnarray*}
Also one has $C\gamma_{5}C^{-1}=\gamma_{5}=\gamma_{5}^{T}$. Then
one finally arrives at \begin{eqnarray*}
\mathcal{T}_{sr,-}^{\lambda\sigma}(-\widehat{\mathbf{p}},-\widehat{\mathbf{k}}) & = & \mathrm{Tr}\left[\gamma^{\lambda T}(1+\gamma_{5})^{T}(\mathcal{P}_{\mathbf{p},s}^{+})^{T}(\Lambda_{\mathbf{p}}^{+}\gamma^{0})^{T}\gamma^{\sigma T}(1+\gamma_{5})^{T}(\mathcal{P}_{\mathbf{k},r}^{+})^{T}(\Lambda_{\mathbf{k}}^{+}\gamma^{0})^{T}\right]\\
 & = & \mathrm{Tr}\left[\gamma^{\lambda}(1-\gamma_{5})\Lambda_{\mathbf{k}}^{+}\gamma^{0}\mathcal{P}_{\mathbf{k},r}^{+}\gamma^{\sigma}(1-\gamma_{5})\Lambda_{\mathbf{p}}^{+}\gamma^{0}\mathcal{P}_{\mathbf{p},r}^{+}\right]^{T}\\
 & = & \mathrm{Tr}\left[\gamma^{\lambda}(1-\gamma_{5})\mathcal{P}_{\mathbf{k},r}^{+}\Lambda_{\mathbf{k}}^{+}\gamma^{0}\gamma^{\sigma}(1-\gamma_{5})\mathcal{P}_{\mathbf{p},s}^{+}\Lambda_{\mathbf{p}}^{+}\gamma^{0}\right]\\
 & = & \mathcal{T}_{rs,+}^{\lambda\sigma}(\widehat{\mathbf{k}},\widehat{\mathbf{p}}),\end{eqnarray*}
where $[\mathcal{P}_{\mathbf{k},r}^{+},\Lambda_{\mathbf{k}}^{+}\gamma^{0}]=0$
has been used.

\section{Calculate quark tensor $\mathcal{T}_{rs,+}^{\lambda\sigma}$ for
spin-one phases}

\label{sec:quark-tensor}The quark tensor $\mathcal{T}_{rs,+}^{\lambda\sigma}$
can be written as a sum of the leading term with the same structure
as in the massless case and the mass correction term, \begin{eqnarray}
\mathcal{T}_{rs,+}^{\lambda\sigma}(\widehat{\mathbf{k}},\widehat{\mathbf{p}}) & = & \omega_{rs}\mathcal{T}^{\lambda\sigma}(K,P)+\delta\mathcal{T}_{rs}^{\lambda\sigma},\label{eq:mass-t}\end{eqnarray}
where $\mathcal{T}^{\lambda\sigma}(K,P)$ is the quark tensor in the
normal phase and given in Eq. (\ref{eq:quark-tensor-nm}). $\delta\mathcal{T}_{rs}^{\lambda\sigma}$
is the correction from the quark mass.

\subsection{Polar phase}

The polar phase is particularly simple. The order parameter or the
condensate has the form \[
\mathcal{M}_{\mathbf{k}}=J_{3}\gamma_{\perp,3}(\widehat{\mathbf{k}}).\]
The projectors $\mathcal{P}_{\mathbf{k},r}^{+}$ do not depend on
the quark momentum $\mathbf{k}$, \begin{eqnarray}
\mathcal{P}_{\mathbf{k},1}^{+} & = & J_{3}^{2},\nonumber \\
\mathcal{P}_{\mathbf{k},2}^{+} & = & 1-J_{3}^{2}.\label{eq:proj-polar}\end{eqnarray}
The trace can be decoupled into a color and a Dirac one. The color
trace is \begin{eqnarray*}
\mathrm{Tr}[\mathcal{P}_{\mathbf{k},1}^{+}\mathcal{P}_{\mathbf{k},1}^{+}] & = & 2,\\
\mathrm{Tr}[\mathcal{P}_{\mathbf{k},2}^{+}\mathcal{P}_{\mathbf{k},2}^{+}] & = & 1,\\
\mathrm{Tr}[\mathcal{P}_{\mathbf{k},1}^{+}\mathcal{P}_{\mathbf{k},2}^{+}] & = & 0.\end{eqnarray*}
The Dirac trace is the same as in the normal phase in Eq. (\ref{eq:quark-tensor-nm}).
So we have \begin{eqnarray*}
 &  & \omega_{11}=2,\;\omega_{22}=1,\;\omega_{12}=\omega_{21}=0,\\
 &  & \delta\mathcal{T}_{rs}^{\lambda\sigma}=0\;(r,s=1,2).\end{eqnarray*}

\subsection{Planar phase}

The order parameter is in the form \begin{eqnarray*}
\mathcal{M}_{\mathbf{k}} & = & J_{1}\gamma_{\perp}^{1}(\widehat{\mathbf{k}})+J_{2}\gamma_{\perp}^{2}(\widehat{\mathbf{k}}).\end{eqnarray*}
The projectors are \begin{eqnarray*}
\mathcal{P}_{\mathbf{k},1}^{+} & = & \frac{1}{1+\widehat{k}_{3}^{2}}\left[J_{1}^{2}(1-\widehat{k}_{1}^{2})+J_{2}^{2}(1-\widehat{k}_{2}^{2})-\{ J_{1},J_{2}\}\widehat{k}_{1}\widehat{k}_{2}+J_{3}\widehat{k}_{3}\gamma_{0}\gamma^{5}\bm{\gamma}\cdot\widehat{\mathbf{k}}\right],\\
\mathcal{P}_{\mathbf{k},2}^{+} & = & 1-\mathcal{P}_{\mathbf{k},1}^{+}.\end{eqnarray*}
We can rewrite $\mathcal{P}_{\mathbf{k},1}^{+}$ and $\mathcal{P}_{\mathbf{k},2}^{+}$
in the form \begin{eqnarray*}
\mathcal{P}_{\mathbf{k},1}^{+} & = & I_{1\mathbf{k}}+I_{2\mathbf{k}}\gamma_{0}\gamma^{5}\bm{\gamma}\cdot\widehat{\mathbf{k}},\\
\mathcal{P}_{\mathbf{k},2}^{+} & = & 1-I_{1\mathbf{k}}-I_{2\mathbf{k}}\gamma_{0}\gamma^{5}\bm{\gamma}\cdot\widehat{\mathbf{k}},\end{eqnarray*}
with \begin{eqnarray*}
I_{1\mathbf{k}} & = & \frac{1}{1+\widehat{k}_{3}^{2}}\left[J_{1}^{2}(1-\widehat{k}_{1}^{2})+J_{2}^{2}(1-\widehat{k}_{2}^{2})-\{ J_{1},J_{2}\}\widehat{k}_{1}\widehat{k}_{2}\right],\\
I_{2\mathbf{k}} & = & \frac{1}{1+\widehat{k}_{3}^{2}}J_{3}\widehat{k}_{3}.\end{eqnarray*}
We obtain \begin{eqnarray*}
\omega_{11} & = & \mathrm{Tr}_{c}\left[I_{1\mathbf{k}}I_{1\mathbf{p}}\right]+\mathrm{Tr}_{c}\left[I_{2\mathbf{k}}I_{2\mathbf{p}}\right],\\
\omega_{22} & = & -1+\mathrm{Tr}_{c}\left[I_{1\mathbf{k}}I_{1\mathbf{p}}\right]+\mathrm{Tr}_{c}\left[I_{2\mathbf{k}}I_{2\mathbf{p}}\right],\\
\omega_{12} & = & \omega_{21}=2-\mathrm{Tr}_{c}\left[I_{1\mathbf{k}}I_{1\mathbf{p}}\right]-\mathrm{Tr}_{c}\left[I_{2\mathbf{k}}I_{2\mathbf{p}}\right],\\
\delta\mathcal{T}_{11}^{\lambda\sigma} & = & \mathrm{Tr}_{c}\left[I_{2\mathbf{k}}I_{2\mathbf{p}}\right]\left[-R_{p}\mathcal{T}^{\lambda\sigma}(K,\widetilde{P}_{0})-R_{k}\mathcal{T}^{\lambda\sigma}(\widetilde{K}_{0},P)+R_{p}R_{k}\mathcal{T}^{\lambda\sigma}(\widetilde{K}_{0},\widetilde{P}_{0})\right],\\
\delta\mathcal{T}_{22}^{\lambda\sigma} & = & \delta\mathcal{T}_{11}^{\lambda\sigma}=-\delta\mathcal{T}_{12}^{\lambda\sigma}=-\delta\mathcal{T}_{21}^{\lambda\sigma},\end{eqnarray*}
where $R_{k}=(E_{k}-k)/k$ , $K=(E_{k},\mathbf{k})$ and $\widetilde{K}_{0}=(k,-\mathbf{k})$.
$\mathcal{T}^{\lambda\sigma}$ is given in Eq. (\ref{eq:quark-tensor-nm}).
In the collinear limit, $\mathbf{k}\parallel\mathbf{p}$ or $\widehat{\mathbf{k}}\approx\widehat{\mathbf{p}}$,
the color traces can be simplified, then we obtain \begin{eqnarray*}
\omega_{11} & = & 2,\;\omega_{22}=1,\;\omega_{12}=\omega_{21}=0,\\
\delta\mathcal{T}_{11}^{\lambda\sigma} & = & \frac{2\widehat{k}_{3}^{2}}{(1+\widehat{k}_{3}^{2})^{2}}\left[-R_{p}\mathcal{T}^{\lambda\sigma}(K,\widetilde{P}_{0})-R_{k}\mathcal{T}^{\lambda\sigma}(\widetilde{K}_{0},P)+R_{p}R_{k}\mathcal{T}^{\lambda\sigma}(\widetilde{K}_{0},\widetilde{P}_{0})\right],\\
\delta\mathcal{T}_{22}^{\lambda\sigma} & = & \delta\mathcal{T}_{11}^{\lambda\sigma}=-\delta\mathcal{T}_{12}^{\lambda\sigma}=-\delta\mathcal{T}_{21}^{\lambda\sigma}.\end{eqnarray*}

\subsection{$A$ phase }

The order parameter is \begin{eqnarray*}
\mathcal{M}_{\mathbf{k}} & = & J_{3}\left[\gamma_{\perp}^{1}(\widehat{\mathbf{k}})+i\gamma_{\perp}^{2}(\widehat{\mathbf{k}})\right].\end{eqnarray*}
The projectors are \begin{eqnarray*}
\mathcal{P}_{\mathbf{k},1}^{+} & = & \frac{1}{2}J_{3}^{2}\left[1+\mathrm{sgn}(\widehat{k}_{3})\gamma_{0}\gamma^{5}\bm{\gamma}\cdot\widehat{\mathbf{k}}\right],\\
\mathcal{P}_{\mathbf{k},2}^{+} & = & \frac{1}{2}J_{3}^{2}\left[1-\mathrm{sgn}(\widehat{k}_{3})\gamma_{0}\gamma^{5}\bm{\gamma}\cdot\widehat{\mathbf{k}}\right],\\
\mathcal{P}_{\mathbf{k},3}^{+} & = & 1-J_{3}^{2},\end{eqnarray*}
where $\mathrm{sgn}(x)$ is the sign of $x$. The results for $\omega_{rs}$
are \begin{eqnarray*}
\omega_{11} & = & \frac{1}{2}[1+\mathrm{sgn}(\widehat{k}_{3})][1+\mathrm{sgn}(\widehat{p}_{3})],\\
\omega_{22} & = & \frac{1}{2}[1-\mathrm{sgn}(\widehat{k}_{3})][1-\mathrm{sgn}(\widehat{p}_{3})],\\
\omega_{33} & = & 1,\\
\omega_{13} & = & \omega_{31}=\omega_{23}=\omega_{32}=0,\\
\omega_{12} & = & \frac{1}{2}[1+\mathrm{sgn}(\widehat{k}_{3})][1-\mathrm{sgn}(\widehat{p}_{3})],\\
\omega_{21} & = & \frac{1}{2}[1-\mathrm{sgn}(\widehat{k}_{3})][1+\mathrm{sgn}(\widehat{p}_{3})].\end{eqnarray*}
The results for $\delta\mathcal{T}_{rs}^{\lambda\sigma}$ are \begin{eqnarray*}
\delta\mathcal{T}_{11}^{\lambda\sigma} & = & -\frac{1}{2}\mathrm{sgn}(\widehat{p}_{3})[1+\mathrm{sgn}(\widehat{k}_{3})]R_{p}\mathcal{T}^{\lambda\sigma}(K,\widetilde{P}_{0})\\
 &  & -\frac{1}{2}\mathrm{sgn}(\widehat{k}_{3})[1+\mathrm{sgn}(\widehat{p}_{3})]R_{k}\mathcal{T}^{\lambda\sigma}(\widetilde{K}_{0},P)\\
 &  & +\frac{1}{2}\mathrm{sgn}(\widehat{p}_{3})\mathrm{sgn}(\widehat{k}_{3})R_{p}R_{k}\mathcal{T}^{\lambda\sigma}(\widetilde{K}_{0},\widetilde{P}_{0}),\\
\delta\mathcal{T}_{22}^{\lambda\sigma} & = & \frac{1}{2}\mathrm{sgn}(\widehat{p}_{3})[1-\mathrm{sgn}(\widehat{k}_{3})]R_{p}\mathcal{T}^{\lambda\sigma}(K,\widetilde{P}_{0})\\
 &  & +\frac{1}{2}\mathrm{sgn}(\widehat{k}_{3})[1-\mathrm{sgn}(\widehat{p}_{3})]R_{k}\mathcal{T}^{\lambda\sigma}(\widetilde{K}_{0},P)\\
 &  & +\frac{1}{2}\mathrm{sgn}(\widehat{p}_{3})\mathrm{sgn}(\widehat{k}_{3})R_{p}R_{k}\mathcal{T}^{\lambda\sigma}(\widetilde{K}_{0},\widetilde{P}_{0}),\\
\delta\mathcal{T}_{33}^{\lambda\sigma} & = & \delta\mathcal{T}_{13}^{\lambda\sigma}=\delta\mathcal{T}_{31}^{\lambda\sigma}=\delta\mathcal{T}_{23}^{\lambda\sigma}=\delta\mathcal{T}_{32}^{\lambda\sigma}=0,\\
\delta\mathcal{T}_{12}^{\lambda\sigma} & = & \frac{1}{2}\mathrm{sgn}(\widehat{p}_{3})[1+\mathrm{sgn}(\widehat{k}_{3})]R_{p}\mathcal{T}^{\lambda\sigma}(K,\widetilde{P}_{0})\\
 &  & -\frac{1}{2}\mathrm{sgn}(\widehat{k}_{3})[1-\mathrm{sgn}(\widehat{p}_{3})]R_{k}\mathcal{T}^{\lambda\sigma}(\widetilde{K}_{0},P)\\
 &  & -\frac{1}{2}\mathrm{sgn}(\widehat{p}_{3})\mathrm{sgn}(\widehat{k}_{3})R_{p}R_{k}\mathcal{T}^{\lambda\sigma}(\widetilde{K}_{0},\widetilde{P}_{0}),\\
\delta\mathcal{T}_{21}^{\lambda\sigma} & = & -\frac{1}{2}\mathrm{sgn}(\widehat{p}_{3})[1-\mathrm{sgn}(\widehat{k}_{3})]R_{p}\mathcal{T}^{\lambda\sigma}(K,\widetilde{P}_{0})\\
 &  & +\frac{1}{2}\mathrm{sgn}(\widehat{k}_{3})[1+\mathrm{sgn}(\widehat{p}_{3})]R_{k}\mathcal{T}^{\lambda\sigma}(\widetilde{K}_{0},P)\\
 &  & -\frac{1}{2}\mathrm{sgn}(\widehat{p}_{3})\mathrm{sgn}(\widehat{k}_{3})R_{p}R_{k}\mathcal{T}^{\lambda\sigma}(\widetilde{K}_{0},\widetilde{P}_{0}).\end{eqnarray*}
In the collinear limit $\widehat{\mathbf{k}}\approx\widehat{\mathbf{p}}$,
we have \begin{eqnarray*}
 &  & \omega_{11}=2\theta(\widehat{k}_{3}),\;\omega_{22}=2\theta(-\widehat{k}_{3}),\;\omega_{33}=1,\\
 &  & \omega_{12}=\omega_{21}=\omega_{13}=\omega_{31}=\omega_{23}=\omega_{32}=0,\end{eqnarray*}
and \begin{eqnarray*}
\delta\mathcal{T}_{11}^{\lambda\sigma} & = & -\theta(\widehat{k}_{3})R_{p}\mathcal{T}^{\lambda\sigma}(K,\widetilde{P}_{0})-\theta(\widehat{k}_{3})R_{k}\mathcal{T}^{\lambda\sigma}(\widetilde{K}_{0},P)+\frac{1}{2}R_{p}R_{k}\mathcal{T}^{\lambda\sigma}(\widetilde{K}_{0},\widetilde{P}_{0}),\\
\delta\mathcal{T}_{22}^{\lambda\sigma} & = & -\theta(-\widehat{k}_{3})R_{p}\mathcal{T}^{\lambda\sigma}(K,\widetilde{P}_{0})-\theta(-\widehat{k}_{3})R_{k}\mathcal{T}^{\lambda\sigma}(\widetilde{K}_{0},P)+\frac{1}{2}R_{p}R_{k}\mathcal{T}^{\lambda\sigma}(\widetilde{K}_{0},\widetilde{P}_{0}),\\
\delta\mathcal{T}_{33}^{\lambda\sigma} & = & \delta\mathcal{T}_{13}^{\lambda\sigma}=\delta\mathcal{T}_{31}^{\lambda\sigma}=\delta\mathcal{T}_{23}^{\lambda\sigma}=\delta\mathcal{T}_{32}^{\lambda\sigma}=0,\\
\delta\mathcal{T}_{12}^{\lambda\sigma} & = & \theta(\widehat{k}_{3})R_{p}\mathcal{T}^{\lambda\sigma}(K,\widetilde{P}_{0})+\theta(-\widehat{k}_{3})R_{k}\mathcal{T}^{\lambda\sigma}(\widetilde{K}_{0},P)-\frac{1}{2}R_{p}R_{k}\mathcal{T}^{\lambda\sigma}(\widetilde{K}_{0},\widetilde{P}_{0}),\\
\delta\mathcal{T}_{21}^{\lambda\sigma} & = & \theta(-\widehat{k}_{3})R_{p}\mathcal{T}^{\lambda\sigma}(K,\widetilde{P}_{0})+\theta(\widehat{k}_{3})R_{k}\mathcal{T}^{\lambda\sigma}(\widetilde{K}_{0},P)-\frac{1}{2}R_{p}R_{k}\mathcal{T}^{\lambda\sigma}(\widetilde{K}_{0},\widetilde{P}_{0}),\end{eqnarray*}
where $\theta(x)\equiv1+\mathrm{sgn}(x)$ is the step function.

\subsection{CSL phase }

The order parameter is \begin{eqnarray*}
\mathcal{M}_{\mathbf{k}} & = & \mathbf{J}\cdot\bm{\gamma}_{\perp}(\widehat{\mathbf{k}}).\end{eqnarray*}
The projectors are \begin{eqnarray*}
\mathcal{P}_{\mathbf{k},1}^{+} & = & -\frac{1}{2}\left[\mathbf{J}\cdot\bm{\gamma}\right]^{2},\\
\mathcal{P}_{\mathbf{k},2}^{+} & = & 1+\frac{1}{2}\left[\mathbf{J}\cdot\bm{\gamma}\right]^{2}.\end{eqnarray*}
It is convenient to rewrite these projectors in terms of color indices,
\begin{eqnarray*}
(\mathcal{P}_{\mathbf{k},1}^{+})_{ab} & = & \delta_{ab}+\frac{1}{2}\gamma_{\bot}^{b}\gamma_{\bot}^{a},\\
(\mathcal{P}_{\mathbf{k},2}^{+})_{ab} & = & -\frac{1}{2}\gamma_{\bot}^{b}\gamma_{\bot}^{a}.\end{eqnarray*}
We obtain \begin{eqnarray*}
\omega_{11} & = & 1+\frac{1}{4}\left[1+(\widehat{\mathbf{k}}\cdot\widehat{\mathbf{p}})^{2}+2(\widehat{\mathbf{k}}\cdot\widehat{\mathbf{p}})\right],\\
\omega_{22} & = & \frac{1}{4}\left[1+(\widehat{\mathbf{k}}\cdot\widehat{\mathbf{p}})^{2}+2(\widehat{\mathbf{k}}\cdot\widehat{\mathbf{p}})\right],\\
\omega_{12} & = & \omega_{21}=1-\frac{1}{4}\left[1+(\widehat{\mathbf{k}}\cdot\widehat{\mathbf{p}})^{2}+2(\widehat{\mathbf{k}}\cdot\widehat{\mathbf{p}})\right],\end{eqnarray*}
and \begin{eqnarray*}
\delta\mathcal{T}_{11}^{\lambda\sigma} & = & \frac{1}{2}(\widehat{\mathbf{k}}\cdot\widehat{\mathbf{p}})\left[-R_{p}\mathcal{T}^{\lambda\sigma}(K,\widetilde{P}_{0})-R_{k}\mathcal{T}^{\lambda\sigma}(\widetilde{K}_{0},P)+R_{p}R_{k}\mathcal{T}^{\lambda\sigma}(\widetilde{K}_{0},\widetilde{P}_{0})\right],\\
\delta\mathcal{T}_{22}^{\lambda\sigma} & = & \delta\mathcal{T}_{11}^{\lambda\sigma}=-\delta\mathcal{T}_{12}^{\lambda\sigma}=-\delta\mathcal{T}_{21}^{\lambda\sigma}.\end{eqnarray*}
In the collinear limit, we have \begin{eqnarray*}
 &  & \omega_{11}=2,\;\omega_{22}=1,\;\omega_{12}=\omega_{21}=0,\\
 &  & \delta\mathcal{T}_{11}^{\lambda\sigma}=\frac{1}{2}\left[-R_{p}\mathcal{T}^{\lambda\sigma}(K,\widetilde{P}_{0})-R_{k}\mathcal{T}^{\lambda\sigma}(\widetilde{K}_{0},P)+R_{p}R_{k}\mathcal{T}^{\lambda\sigma}(\widetilde{K}_{0},\widetilde{P}_{0})\right],\\
 &  & \delta\mathcal{T}_{22}^{\lambda\sigma}=\delta\mathcal{T}_{11}^{\lambda\sigma}=-\delta\mathcal{T}_{12}^{\lambda\sigma}=-\delta\mathcal{T}_{21}^{\lambda\sigma}.\end{eqnarray*}

\end{document}